\documentclass[%
 reprint,
 amsmath,amssymb,
 aps,
 prl
]{revtex4-1}

\usepackage{graphicx}% Include figure files
\graphicspath{{./figures/}}
\usepackage{dcolumn}% Align table columns on decimal point
\usepackage{bm}% bold math

\usepackage{float}
\usepackage[usenames]{color}

\usepackage{ulem}

\begin{document}

\preprint{APS/123-QED}

\title{The equilibrium structure of self-assembled protein nano-cages}
\author{Sanaz Panahandeh}
 
\author{Siyu Li}%
 \author{Roya Zandi}%
\affiliation{%
 Department of Physics and Astronomy, University of California, Riverside, California 92521, USA
}%


%\affiliation{
%\affiliation{
 %Third institution, the second for Charlie Author
%\affiliation{%



\begin{abstract}
Understanding how highly symmetric, robust, monodisperse protein cages self-assemble can have major applications in various areas of bio-nanotechnology, such as drug delivery, biomedical imaging and gene therapy.   We develop a model to investigate the assembly of protein subunits into the structures with different size and symmetry.  Using Monte Carlo simulation, we obtain the global minimum energy structures. Our results suggest that the physical properties of building blocks including the spontaneous curvature, flexibility and bending rigidity of coat proteins are sufficient to predict the size of the assembly products and that the symmetry and shape selectivity of nano-cages can be explained, at least in part, on a thermodynamic basis.   The polymorphism of nano-cages observed in {\it vitro} assembly experiments are also discussed.

% \begin{description}
% \item[PACS numbers]
% May be entered using the \verb+\pacs{#1}+ command.
% \end{description}
\end{abstract}

% \pacs{Valid PACS appear here}% PACS, the Physics and Astronomy
\maketitle

%\tableofcontents

\section{\label{sec:level1}Introduction}

Self-assembly of monodispersed protein cages is ubiquitous in nature. Because of their biocompatibility, stability and low toxicity, protein cages have important roles in many biological processes, medicine and bio-nanotechnology. Examples of protein cages include platonic hydrocarbons, heat shock proteins, ferritins, carboxysomes, silicages, multicomponent ligand assemblies, clathrin vesicles and virus shells, to name a few \cite{kroto1985c60, baughman1993fullereneynes, aumiller2018protein, fontana2014phage, ma2018self}.  The protein shells are necessary for both protection and delivery of various cargos in biological systems. For instance, ferritin stores iron and exists in almost every living organism.  

Among all biological entities, viruses in particular have optimized the feat of packaging of genetic materials and other anionic cargos into a protein shell called the capsid, recognized as one of the most efficient nano-containers for trafficking genetic material in nature \cite{elife,Gonca2016}. Most protein cages self-assemble from a large number of one or a few different types of protein subunits into complex supramolecular structures with diameters ranging from 10 to 500 nm \cite{heddle2008protein}.   Quite remarkably under many circumstances, viruses spontaneously assemble {\it in vitro} from protein building blocks into highly symmetric shells \cite{Garmann2015, chevreuil2018nonequilibrium}. Most spherical viruses adopt structures with icosahedral symmetry \cite{crick1956structure, weiss2005armor, johnson1997quasi} characterized by a structural index $T$ number, which assume only certain integers (1, 3, 4, 7, ...) \cite{CASPAR1962}. The number of protein subunits in icosahedral shells is often $60$ times the $T$-number.

\begin{figure}

\centering
\includegraphics[width=0.5\textwidth]{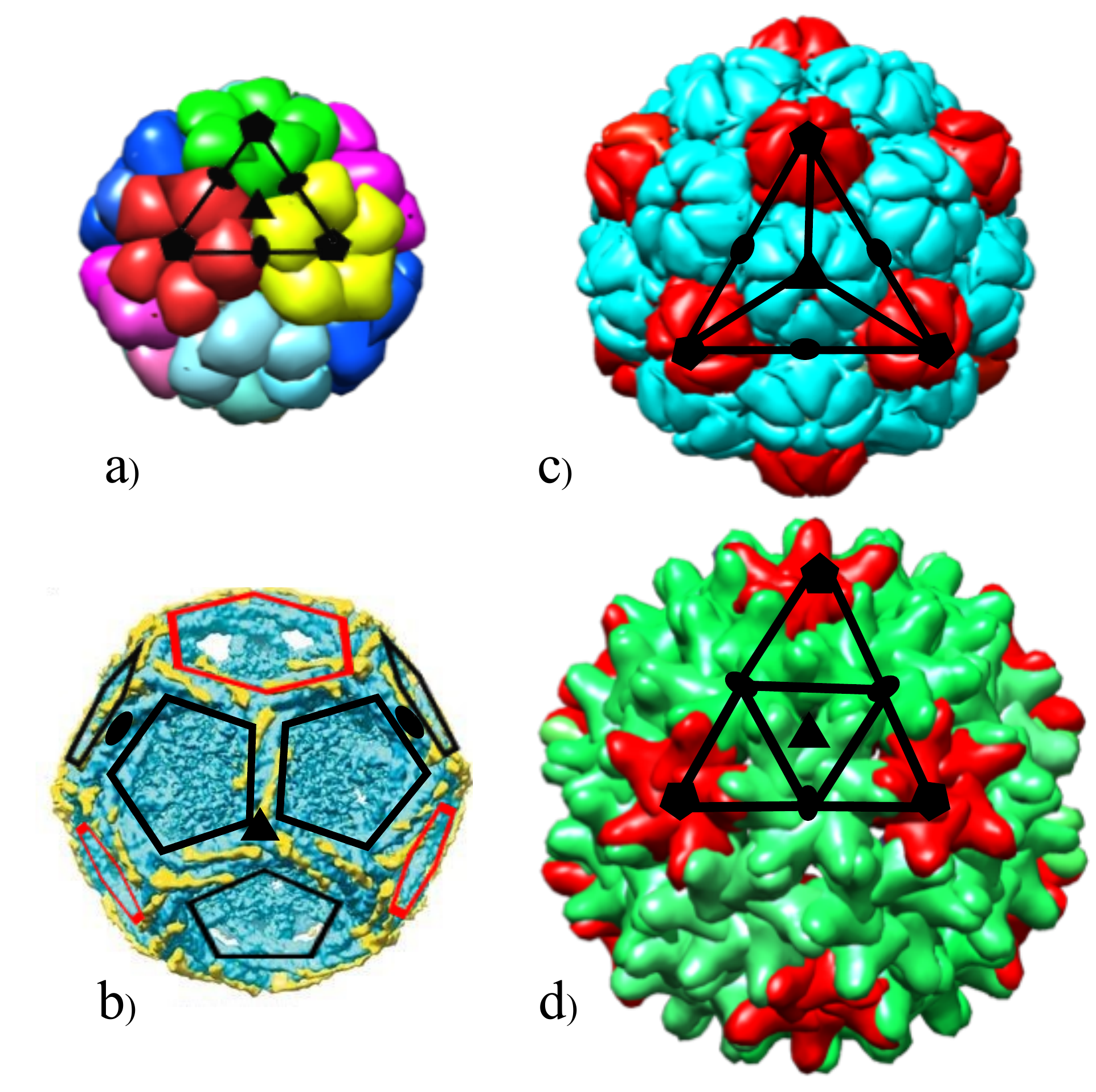}
\caption{Structures of some protein cages. (a) Lumazine synthase is an enzyme with icosahedral symmetry ($T=1$) constructed of 60 identical protein subunits. The colors are added to highlight each pentamer. (b) Mini-coat has tetrahedral symmetry \cite{fotin2004molecular}. The two fold and tree fold symmetry axis are marked with small black ovals and triangles respectively. (c) Encapsulin from M. xanthus with $T=3$ structure is made of 180 identical protein subunits. The position of two, three and five fold symmetry axis are marked in the picture.  (d) Hepatitis B virus, a $T=4$ structure. The darker color (red) in (c) and (d) are pentamers. All the structures except (b) are reproduced using UCSF Chimera packages (http://www.rbvi.ucsf.edu/chimera).} 
%Chimera is developed by the Resource for Biocomputing, Visualization, and Informatics at the University of California, San Francisco (supported by NIGMS P41-GM103311).}
\label{examples}
\end{figure}

Other protein cages can adopt several other symmetric structures. For example, clathrin shells form icosahedral structures in addition to many other symmetric shells \cite{cheng2007cryo, crowther1976structure} depending on the size of their cargo. Nevertheless protein cages with icosahedral symmetry are by far the most abundant in nature. Figure \ref{examples} illustrates the structure of Lumazine synthase with $T=1$ symmetry\cite{min2014lumazine}, a Clathrin shell with tetrahedral symmetry \cite{fotin2004molecular}, an encapsulin nanocompartment from M. xanthus with $T=3$ structure \cite{mchugh2014virus} and the Hepatitis B virus (HBV) capsid  with $T=4$ symmetry. 

Despite the abundance of protein shells in nature, the role of building blocks and the factors contributing to the stability, size and shape selectivity of nanostructures are not well-understood.  To this end, there is a precedent need to take a bottom-up approach and to understand at the fundamental scale the impact of building blocks on the design and formation of functional nano-shells.

Extensive work has explored the effect of spontaneous radius of curvature (dihedral angle) of building blocks on the equilibrium structure of protein cages \cite{Zandi2016}. For instance, using the Monte Carlo (MC) simulations, Chen {\it et al.} studied the self-assembly of attractive cone-shaped particles into different structures \cite{doi:10.1021/la063755d}.  They obtained a sequence of clusters and found that the symmetry and stability of formed structures depend on the cone angle or the preferred angle between subunits. Similar sequence of structures was obtained with attractive spherical particles but under certain convexity constraints, equivalent of changing the preferred dihedral angle between subunits \cite{Chen:2007b}.

The simple case of $N$ spherical colloids or circular disks interacting through Lennard-Jones potential constrained to move on the surface of a sphere also shows that the equilibrium structure of shells depends on the number of building blocks and the preferred angle between disks or Lennard-Jones particles \cite{zandi2004}. As the preferred angle between disks or colloids changes, structures with different size and symmetries form.

More recently, Paquay {\it et al.} studied the equilibrium structures of interacting Morse particles residing on the surface of a sphere and found similar structures and magic numbers as observed in the case of LJ particles \cite{Stefan}. Nevertheless, the impact on the equilibrium structures of the mechanical properties of building blocks including flexibility and bending rigidity have not previously been studied.  While the dynamical structures of protein shells under non-equilibrium conditions as a function of bending rigidity and stretching modulus of building blocks have been thoroughly investigated in Ref. \cite{Wagner2015}, due to irreversible steps in the shell growth, the structures obtained in those simulations might be completely far from equilibrium. 

In this paper we investigate the equilibrium structure of nano-shells and the important factors contributing to their stability and symmetry. Using MC simulations combined with the bond flipping method \cite{kohyama2003budding, rotskoff2018robust} we study the structure of protein cages as a function of the spontaneous curvature as well as stretching and bending rigidity of building blocks, advancing our knowledge for producing high yield nano-cages with specific size and shape.

While spontaneous curvature is an important factor in defining the size of the shell, we find that the flexibility and bending rigidity of building blocks can completely modify the size and final symmetry of the shells.  Quite interestingly the sequence of clusters or magic numbers and their associated shells obtained in our equilibrium studies, coincide not only with the structure of viruses displaying icosahedral symmetry but with other non-icosahedral protein cages observed in other systems such as clathrin shells. 

Finally, we find that there are striking similarities between equilibrium and non-equilibrium ``shape'' phase diagrams as a function of the mechanical properties and spontaneous curvature of building blocks. The fact that pentamers form in the ``correct'' positions during the assembly process of symmetric shells, even under non-equilibrium conditions is quite unexpected.  This reveals that there is a large affinity for the formation of disclinations (pentamers) at {\it specific} locations during the growth of most symmetric nano-cages, which are built from subunits that can form hexagonal sheets in flat space. 
Nevertheless, we find some differences between two phase diagrams too. A few symmetric structures grown in irreversible simulations \cite{Wagner2015} do not constitute the minimum free energy structures. Furthermore, we obtain additional symmetric structures in the equilibrium simulations, which were not observed in the growth simulations under non-equilibrium conditions.   

It is worth mentioning that it is now widely accepted that the preferred curvature and mechanical properties of subunits depend on the solution conditions such as pH and salt concentration \cite{Fejer:10}. The interplay of protein geometry, repulsive electrostatic and attractive hydrophobic interactions define the equilibrium properties (bending and stretching moduli and spontaneous curvature) of subunits, nevertheless no systematic experimental data are known for these parameters.  In addition to solution conditions, mutations can also affect the physical properties of protein subunits \cite{Venky2016}, enabling us to test several theories in this paper.  Understanding the role of stiffness and preferred curvature of building blocks could lead to generation of a range of new materials and novel structures.  

%The article is organized as follows. In the next section, we present our model and describe the methods for obtaining the equilibrium structures. In Sect. III we present our results corresponding to the equilibrium structure of nanocages. We next compare the equilibrium simulated structures with those obtained under non-equilibrium conditions in Sect. III. A. Finally, in Sect. V. we discuss the implications of our simulation results in the context of current understanding of the assembly of protein cages and summarize our findings. 

% Fig. 1

\section{\label{sec:level2}Method}

To study the equilibrium structures, we consider stretchable equilateral triangular subunits, representing building blocks of protein cages, as  illustrated in Fig.~\ref{examples}a and d. Triangular subunits are suitable to describe the structure of protein cages as they form hexagonal sheets in flat space and are able to build a spherical mesh with at least 12 five coordinated lattice points (pentagons).  The total energy of a triangular shell is the sum of the stretching and bending energies \cite{PhysRevE.74.031912}. The stretching energy of triangular network can simply be defined by a harmonic potential summed over all triangles,

\begin{equation}
E_s=\sum_{i} \sum_{a=1}^{3} \frac{k_s}{2}\ (b_{i}^a-b_{0})^2
\label{stretching energy}
\end{equation}
with $i$ the triangular subunit index, $b_0$ the equilibrium length of the edges, and $b_{i}^a$ the length of the $a^{th}$ edge in the $i^{th}$ subunit. While the stretching energy is related to the deformation of subunits from their equilateral shapes, the bending energy corresponds to the deviation of the dihedral angle between adjacent subunits from the preferred one. The bending energy is obtained by summing over all pairs of triangular subunits that share an edge and can be written as,

\begin{equation}
E_b=\sum_{<ij>} k_b (1-\cos(\theta_{ij}-\theta_{0}))
\label{bending energy}
\end{equation}
with $\ <ij>$ the index pairs of neighboring subunits, $k_b$ the torsional spring constant, and $\theta_0$ the preferred dihedral angle between two subunits.  The preferred dihedral angle and radius of curvature are related through $\sin(\theta_0/2)=\ (12R_{0}^2/b_{0}^2\ -3)^{-1/2}$ with $R_0$ the spontaneous radius of curvature. The angle $\theta_{ij}$ is between the unit normal vectors $\hat{n_i}$ and $\hat{n_j}$ of the two adjacent subunits $i$ and $j$ ($\cos\theta_{ij}=\hat{n_i}\cdot\hat{n_j}$), sharing an edge.

Equations \ref{stretching energy} and \ref{bending energy} reveal the presence of two important dimensionless parameters, the spontaneous radius of curvature $R_0/b_0$ and the Foppl von Karman (FvK) number
\begin{equation}
\gamma = k_s b_0^2/k_b,
\label{gamma}
\end{equation}
which indicates the relative difficulty of deforming an equilateral triangular subunit compared to changing the dihedral angle between two adjacent subunits away from the preferred one. We note that both dimensionless parameters are normalized with respect to the size of the subunits $b_0$.

\begin{figure}

\centering
\includegraphics[width=0.5\textwidth]{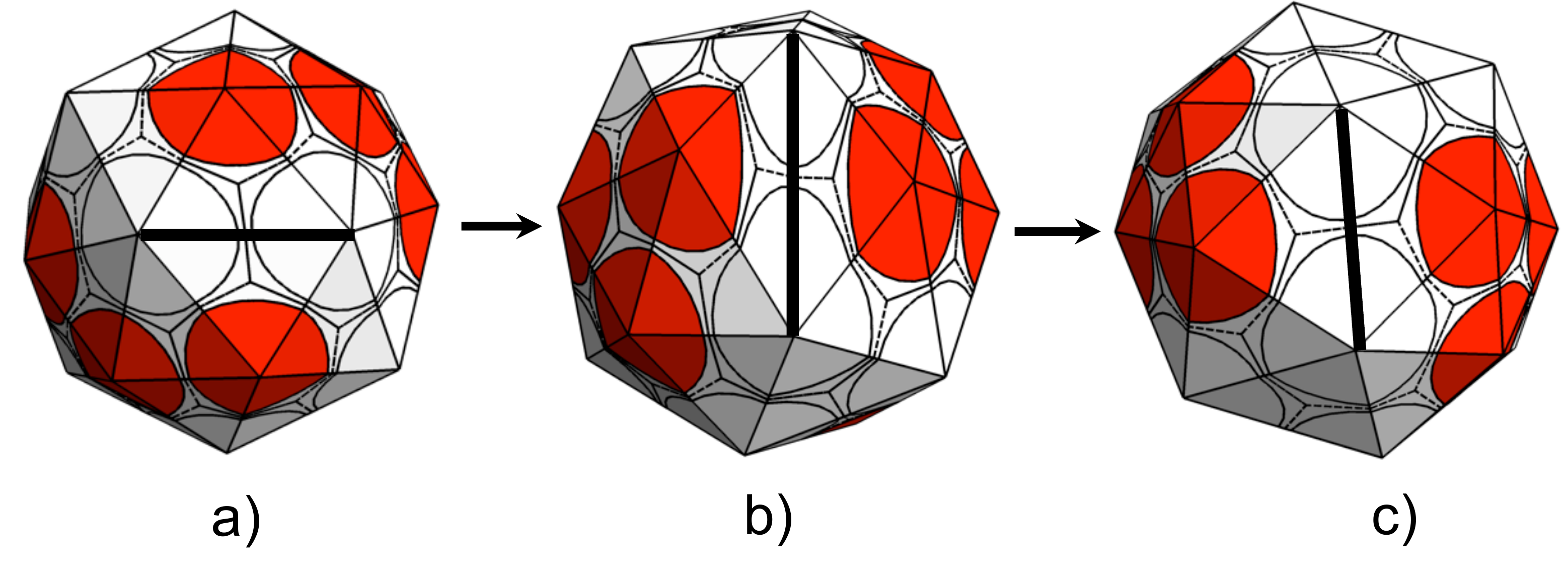}
\caption{Bond moving method: (a) The thick black edge between two neighboring triangles is randomly chosen. (b) The black line is removed from its previous position and the two vertices that were not sharing a bond before the swap, are now connected.  The darker (red) shades indicate the positions of pentamers and the white ones correspond to hexamers. By moving the bond from (a) to (b), the position of pentamers and hexamers are changed. (c) The system is energetically relaxed now after the swap.}
\label{bondchange}
\end{figure}

To obtain the lowest-energy configurations we employ a series of simulated annealing MC simulations \cite{Li6611}.   We start from a triangulated spherical mesh with a random distribution of $N_v$ vertices. Each MC step consists of $N_v$   attempted edge swaps, which involves removing and reattaching the edge connecting two vertices of two neighboring triangles such that the two vertices which were not connected before,  they will be linked by an edge after the flip, as shown in Fig.~\ref{bondchange}. Each edge swapping is followed by the shell relaxation during which vertices will move to the positions that minimize the total elastic energy. We employ the BFGS method to relax and minimize the energy of the shell \cite{NoceWrig06}.  The probability that the new relaxed structure with the new position of vertices to be accepted is $\min(1, e^{(E_{old}-E_{new})/k_BT})$. $E_{old}$ and $E_{new}$ are the energies of the structures before and after the trial edge swapping, respectively. Since our goal is to obtain the global minimum energy structure at low temperature, we relax the system after each edge swapping. We generate a Markov chain with Boltzmann probabilities by iterating the edge swapping until the energy converges. The edge swapping process is reversible to ensure detailed balance.  

We repeat the above simulations with different initial configurations many times.  We also employ different cooling paths to avoid local minimum free energy traps for achieving equilibrium structures. Figure \ref{bondchange} illustrates a bond flip move. The thick edge in Fig.~\ref{bondchange}a is removed but then it will be linked to two other vertices in Fig.~\ref{bondchange}b. After that we allow the shell to relax as illustrated in Fig.~\ref{bondchange}c.  Detachment and reconnection of the bonds are such that the total number of vertices $N_v$, subunits $n_s$ and edges in the shell remain constant (Fig. \ref{bondchange}).  

This algorithm allows us to successfully change the position of pentamers and hexamers. In other words, during the simulations the location of disclinations is not fixed; they can move, and thus change the structure and symmetry of the shell. We perform MC simulations for all the structures ranging from $N_v=12$ to $42$ corresponding to the shells made of $n_s=20$ to $80$ number of subunits.   Since in nature larger shells need some external help like scaffolding proteins or inner core to form symmetric shells \cite{dokland1999scaffolding}, the focus of this paper is on the smaller shells that are able to assemble spontaneously without any core.  

% fig 2

\section{\label{sec:level3}Results}

\begin{figure}

\centering
\includegraphics[width=0.5\textwidth]{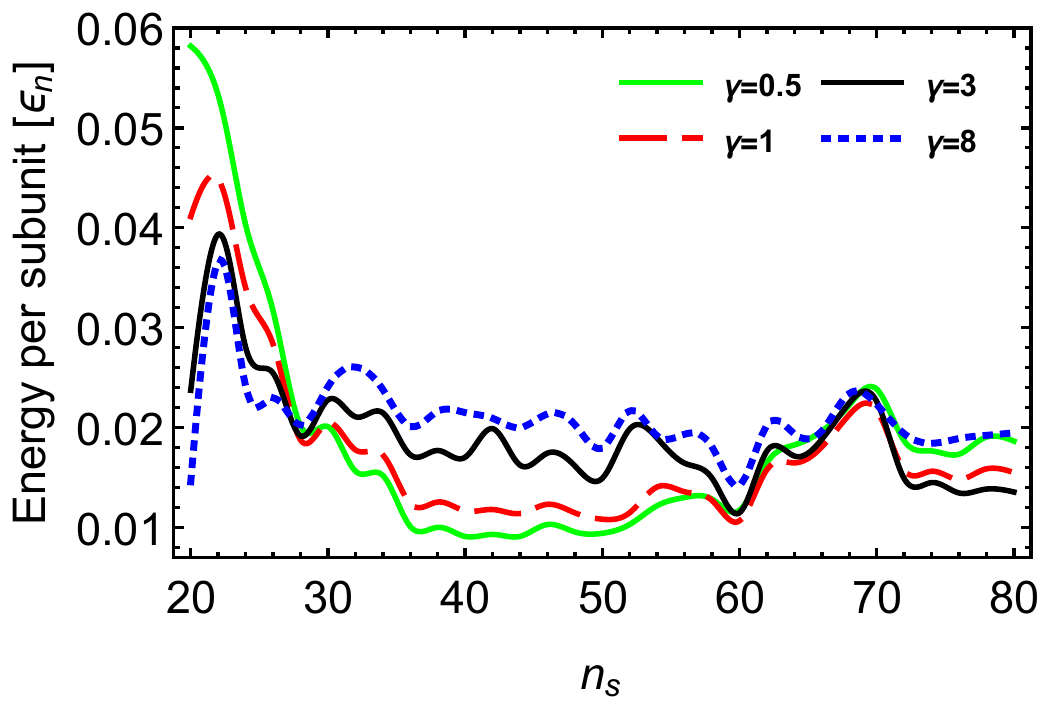}
\caption{Energy per subunit versus number of subunits are displayed for $\gamma=0.5,~1,~3$ and $8$. The spontaneous curvature is fixed at  $R_0/b_0=1.28$.  While the minimum energy (equilibrium structure) for $R_0/b_0=1.28$ and $\gamma=0.5$ is at $n_s=40$, for $\gamma=1,~3$ and $8$ the equilibrium structure is $T=3$ icosahedral structure.}
\label{Energy vs Nv}
\end{figure}

We carry out Monte Carlo simulations as described in the previous section for different number of subunits $n_s$. We start with a fixed preferred spontaneous radius of curvature $R_0/b_0=1.28$ but different values of $\gamma$. The results of the simulations are illustrated in Fig.~\ref{Energy vs Nv} in the form of a plot of the minimized elastic energy per triangles $\epsilon_n$ (in units of $k_BT$) versus the number of subunits, $n_s$.  The solid light line (green) corresponds to $\gamma=0.5$, the dashed line to $\gamma=1$, the dark solid line  to $\gamma=3$ and the dotted line to $\gamma=8$.  We emphasize once more that since $\gamma$ is proportional to the ratio of stretching to bending modulus, for larger $\gamma$s it is difficult to deform the subunits from their equilibrium equilateral shape but rather easy to bend them away from their preferred dihedral angle. For small $\gamma$s, in contrast, the subunits can be easily deformed but it costs significantly more energy to modify the dihedral angle between the adjacent subunits from the preferred one.  

Figure~\ref{Energy vs Nv} illustrates that there are many local minima but no distinguished global minimum energy structure for $\gamma=0.5$. However as $\gamma$ increases the local energy minima corresponding to certain structures become deeper and more pronounced and the $T=3$ icosahedral shell becomes a global minimum for $\gamma \geqslant 1$. For $\gamma<1$, other structures with different $n_s$ compete or have lower energies than a $T=3$ structure.  

\begin{figure}

\centering
\includegraphics[width=0.5\textwidth]{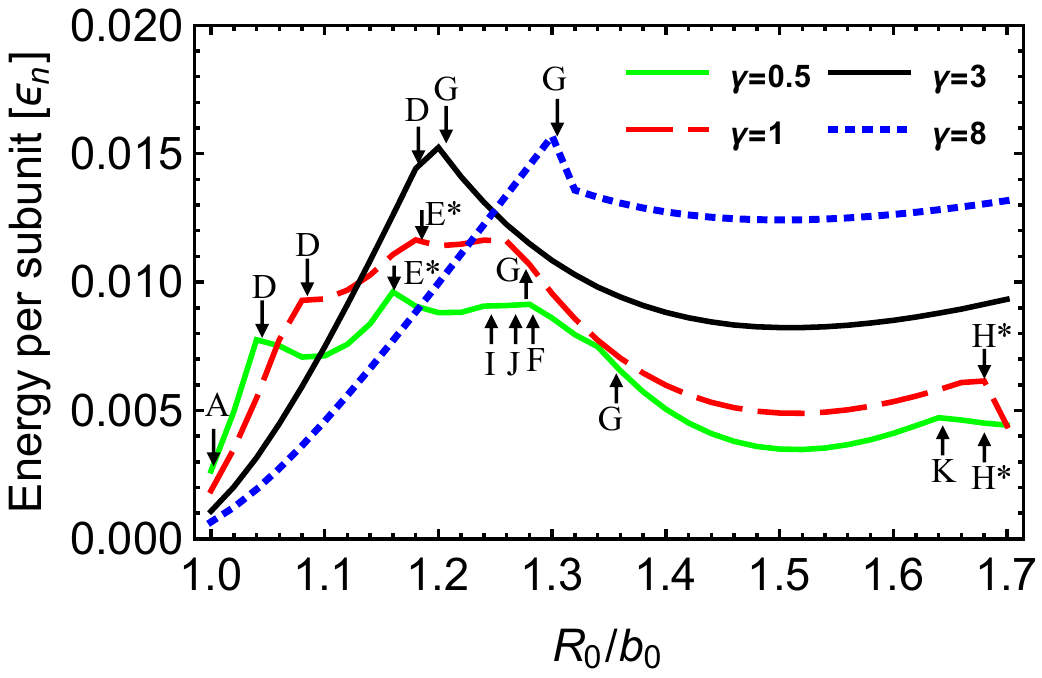}
\caption{Plot of energy per subunit versus $R_0/b_0$ for $\gamma=0.5,~1,~3$ and $8$. Each curve can be divided into different segments. The capital letter at the beginning of each segment (from left to right) indicate the symmetry of the segment. The corresponding structures are illustrated in Fig. \ref{structures}.}
\label{E-R0}
\end{figure}

\begin{figure}

\centering
\includegraphics[width=0.5\textwidth]{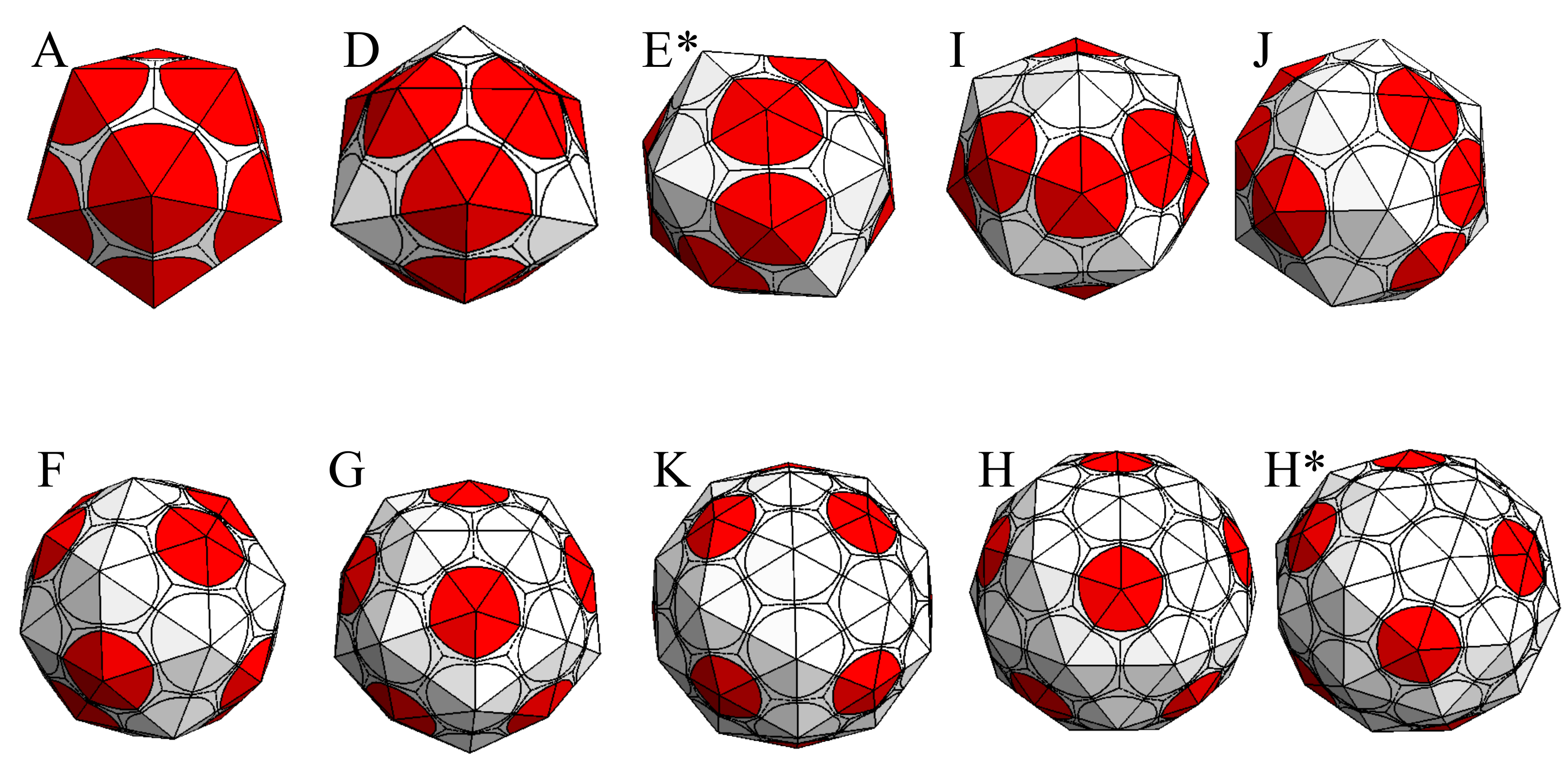}
\caption{The equilibrium structures obtained in the simulations corresponding to the labeled regions of the phase diagram illustrated in Fig. \ref{equil-phase}. The shells from left to right and top to bottom have $n_s=20,~28,~36,~40,~44,~50,~60,~76,~80$ and $80$ subunits and symmetries are icosahedral ($T=1$), tetrahedral, $D_2$(tennis ball), $D_2$, $D_2$, $D_3$, icosahedral ($T=3$), tetrahedral, icosahedral ($T=4$) and $D_{5h}$, respectively.}
\label{structures}
\end{figure}

We next investigate the impact of the spontaneous radius of curvature on the global energy minima of Fig.~\ref{Energy vs Nv} for various $\gamma$s.  Figure \ref{E-R0} illustrates the plot of energy per subunit versus $R_0/b_0$ for the global minimum energy structures (Fig.~\ref{Energy vs Nv}) at different $\gamma$-values.  The curves in Fig.~\ref{E-R0} can be divided into different segments, each representing different structure.  The capital letter at the beginning of each segment reveals the symmetry and structure of that segment.  For instance, the letter $A$ at the beginning of the dotted line shows that for $\gamma=8$ the global minimum energy structure is a $T=1$ icosahedral shell when $1<R_0/b_0<1.3$. Note that all the structures corresponding to the capital letters are illustrated in Fig.~\ref{structures}.  

\begin{figure}

\centering
\includegraphics[width=0.5\textwidth]{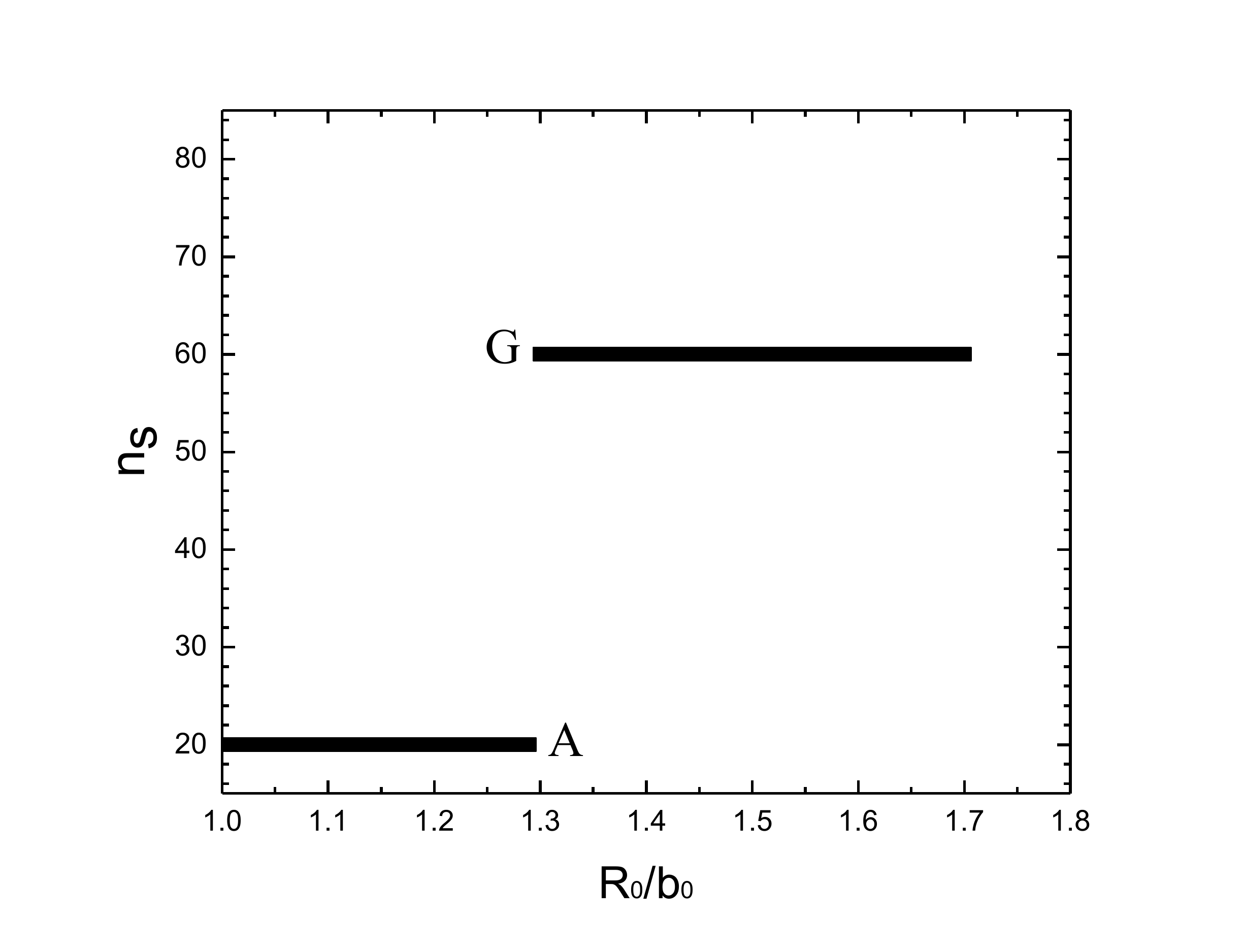}
\caption{Plot of number of subunits in the equilibrium structures versus the spontaneous radius of curvature $R_0/b_0$ at $\gamma=8$. The two flat lines in the plot correspond to $T=1$ and  $T=3$ icosahedral structures.}
\label{N-R0-g8}
\end{figure}

The dotted line in Fig.~\ref{E-R0} shows that even though the energy per subunit increases as $R_0/b_0$ increases, $T=1$ remains the global minimum energy structure till $R_0/b_0=1.3$ when the icosahedral $T=3$ becomes the global minimum energy structure.  This effect is more apparent in Fig.~\ref{N-R0-g8}, which is a plot of number of subunits $n_s$ versus $R_0/b_0$. There is a big jump in the number of subunits from $n_s=20$ ($T=1$) to $n_s=60$ ($T=3$) at $R_0/b_0=1.3$. 

\begin{figure}

\centering
\includegraphics[width=0.5\textwidth]{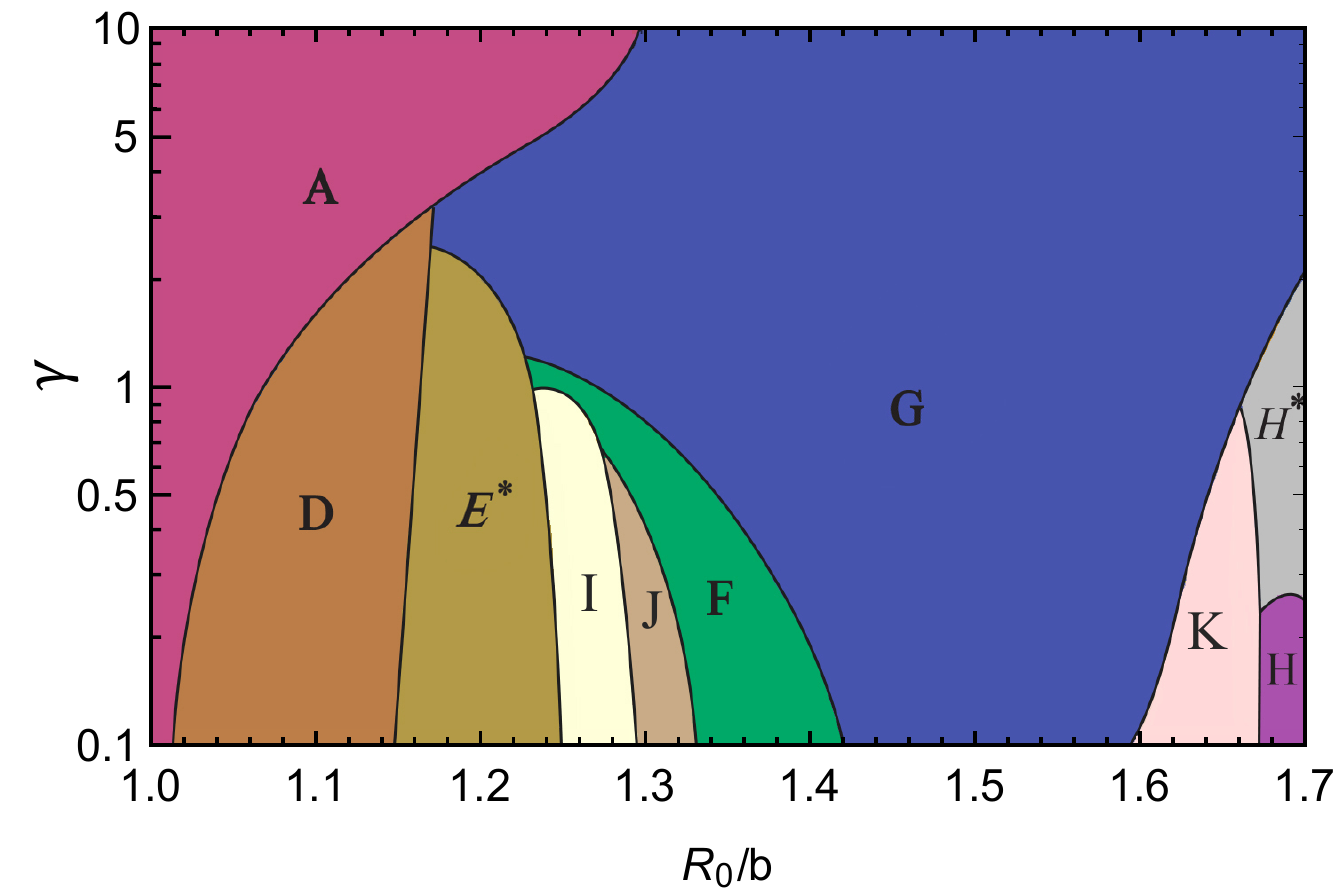}
\caption{Phase diagram of the equilibrium structures presenting various shells assembled for different values of $\gamma$ and $R_0/b$. Each shaded region corresponds to a single equilibrium shell type . Region ($A$) corresponds to a shell with $n_s=20$. The regions $D-K$ correspond to shells with $n_s=28,~36,~50,~60,~80,~40,~44$ and $76$ subunits. Each shell with its corresponding symmetry is shown in Fig. \ref{structures}. Both $H$ and $H^*$ structures have the same number of subunits $n_s=80$.}
\label{equil-phase}
\end{figure}

\begin{figure}

\centering
\includegraphics[width=0.5\textwidth]{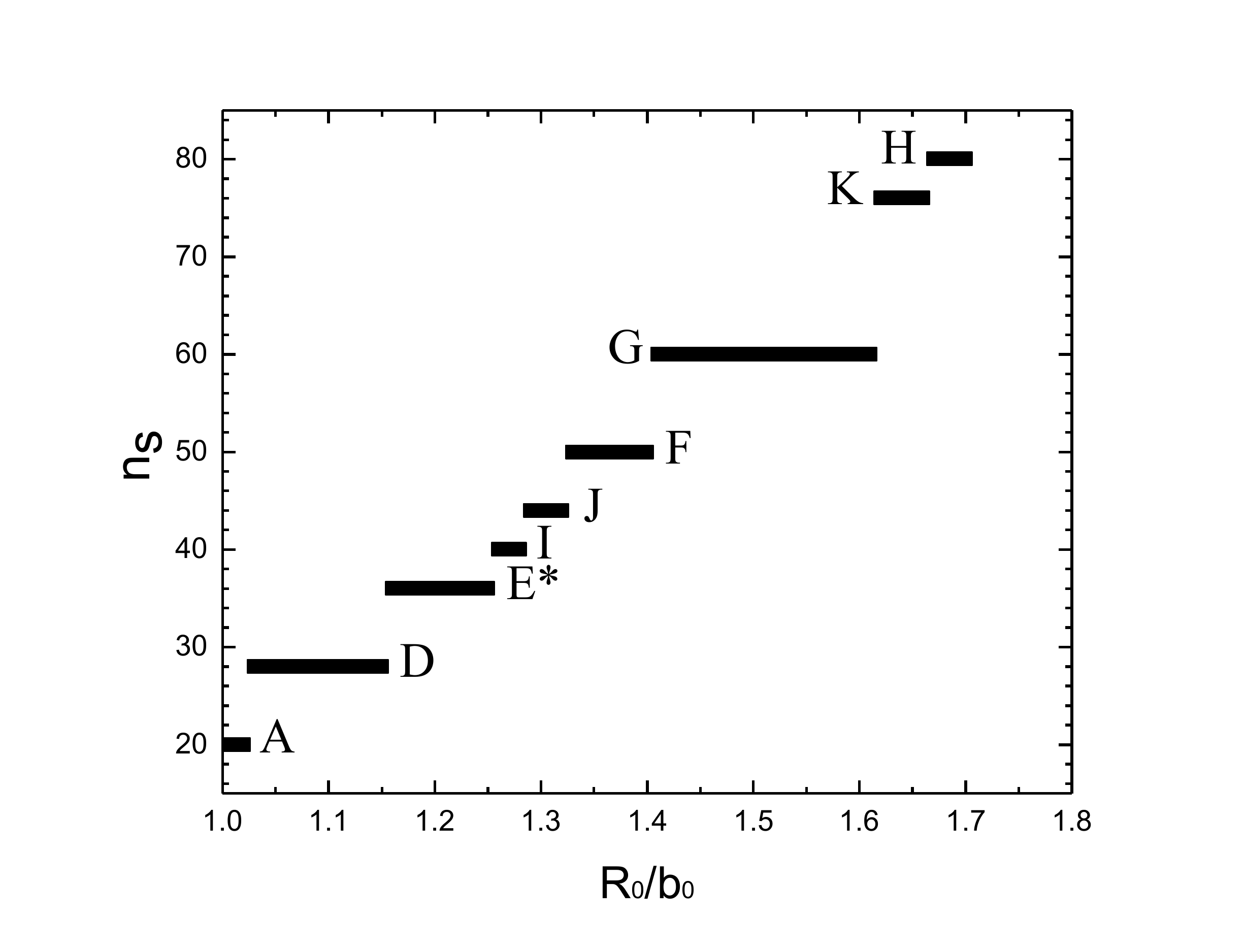}
\caption{Number of subunits in the equilibrium structures versus the spontaneous radius of curvature $R_0/b$ at $\gamma=0.2$. The label next to each line shows the associated structure, illustrated in Fig. \ref{structures}.}
\label{N-R0-g0.2}
\end{figure}

All the above effects can be seen more clearly in Fig.~\ref{equil-phase} in the form of a ``shape'' phase diagram of spontaneous radius of curvature $R_0/b_0$ and $\gamma$. Each shaded region in the diagram corresponds to a different shell whose structure and symmetry are illustrated in Fig.~\ref{structures}.  Figure \ref{equil-phase} shows that the structures become more sensitive to the spontaneous radius of curvature as $\gamma$ decreases. For instance, for $\gamma=0.2$ as the spontaneous radius of curvature varies, we obtain nine different symmetric shells between $R_0/b_0=1$ and $1.7$, see also Fig.~\ref{N-R0-g0.2}. However, there are only two different structures at $\gamma=8$ over a wide range of spontaneous curvature, $T=1$ and $T=3$. This is basically due to the fact that at larger $\gamma$s the protein building blocks are stiffer and it is energetically more costly to deform them from their native shape.  Since for icosahedral structures most proteins are sitting in equivalent positions, at high $\gamma$-values icosahedral structures are the minimum energy structures for the range of the spontaneous curvature studied, as illustrated in Fig.~\ref{equil-phase}.  

The largest shell obtained in Fig.~\ref{equil-phase} contains $n_s=80$ triangles corresponding to a $T=4$ structure for smaller $\gamma$-values.  Note that at intermediate $\gamma$s, another equilibrium structure with the same number of subunits as $T=4$ shell ($n_s=80$) but different symmetry exists, which we label it as $H^*$ in Figs.~\ref{structures} and \ref{equil-phase}.   While $T=1$ and $T=3$ occupy large regions in the equilibrium phase diagram, only a small region belongs to $T=4$.   This is consistent with all previous studies. First of all, a review of literature shows that there are fewer $T=4$ structure in nature \cite{carrillo2008viperdb2, hulo2010viralzone}.   Furthermore, in Refs. \cite{Chen:2007b, Stefan}  only the $D_{5h}$ structure mentioned above was observed and no $T=4$ icosahedral structures appeared in their simulations.

It is now interesting to compare the equilibrium ``shape'' phase diagram with the diagram obtained through irreversible assembly \cite{Wagner2015}.

% fig 3

% fig 4

% fig 5

% fig 6

% fig 7

% fig 8

%
%% fig 6
%\begin{figure}
%
%\centering
%\includegraphics[width=0.5\textwidth]{{ns-R0-g1}.pdf}
%\caption{Number of subunits in the equilibrium structures as a function of spontaneous radius of curvature at $\gamma= 1$}
%\label{N-R0-g1}
%\end{figure}

\subsection{\label{sec:level4}Equilibrium versus non-equilibrium}

\begin{figure}

\centering
\includegraphics[width=0.5\textwidth]{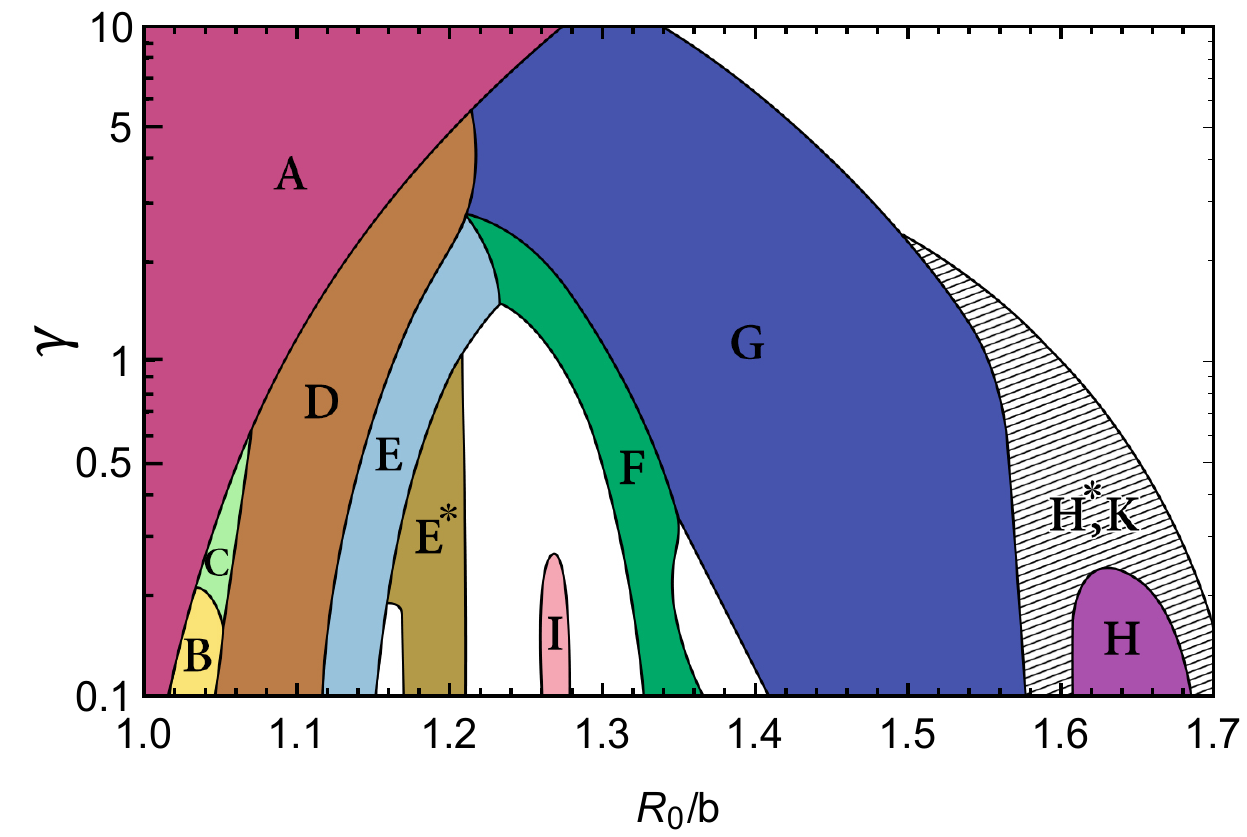}
\caption{The dark shaded contiguous regions labeled by letters ($A$) through ($H$) correspond to regions where only a single type of symmetric shell is assembled. In the hashed region some irregular shells grow in addition to ($H^*$) and ($K$) structures. The majority of $K$ structures form at the phase boundary between $H^*$ and the adjacent white region. The white areas show the regions in which different types of shells without any specific symmetry are formed.}
\label{non-equil-phase}
\end{figure}

  The structures obtained through the irreversible pathway are illustrated in the form of a phase diagram of the dimensionless ratio of bending to stretching modulus ($\gamma$) and the spontaneous radius of curvature in Fig.~\ref{non-equil-phase}.  As in the case of equilibrium phase diagram, each color refers to a different symmetric structure.  While the shells in Fig.~\ref{non-equil-phase} grow following the local minimum free energy pathway, during the assembly process once a pentamer or hexamer forms, its position is permanently fixed. Thus the structures of assembled shells could be completely far from equilibrium. 

We find it quite striking that the shell assembly along the local minimum free energy path with the restrictive conditions of irreversible growth leads to the formation of shells almost identical to those obtained in equilibrium studies.  These results are quite unexpected considering that the principles of detailed balance is violated in the irreversible growth and as such one would expect a big difference between the two phase diagrams.

\begin{figure}

\centering
\includegraphics[width=0.5\textwidth]{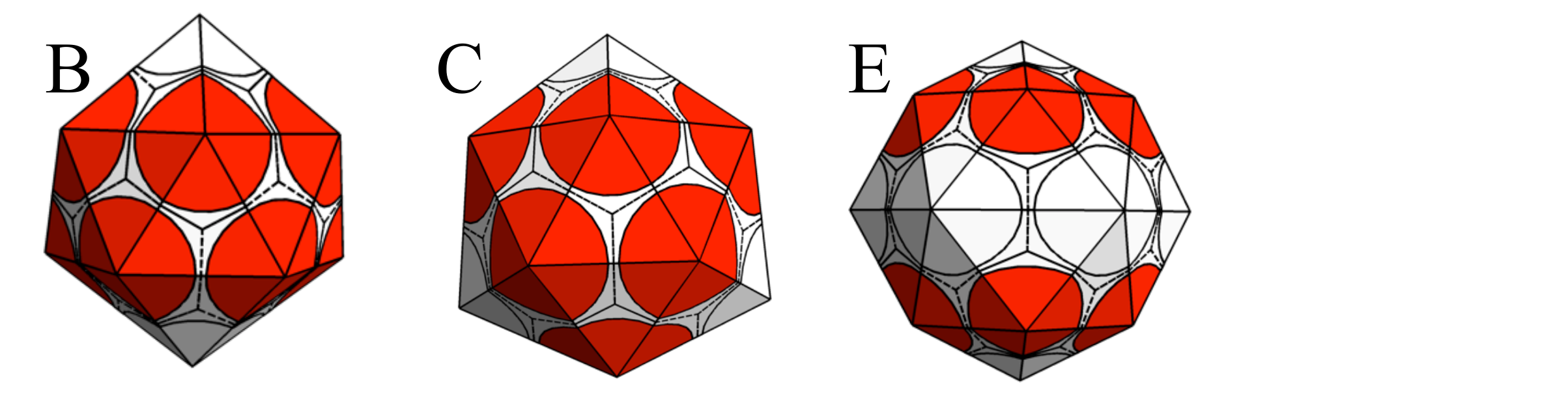}
\caption{The symmetric shells growing under non-equilibrium assembly conditions, which do not appear in the equilibrium phase diagram. The shells from left to right have $n_s=24,~26$ and $36$ subunits. The symmetries are $D_6$, $D_3$ and $D_{6h}$ respectively. In the non-equilibrium phase diagram (Fig. \ref{non-equil-phase}) they are labeled as $B,~C$ and $E$. The $E$ structure has the same number of subunits as $E^*$ structure but with different symmetry.}
\label{structures2}
\end{figure}

Despite the similarities, there are some differences between the two phase diagrams, see Figs.~\ref{equil-phase} and \ref{non-equil-phase}. The shells that only appear in the non-equilibrium phase diagram are illustrated in Fig.~\ref{structures2}.   Two small regions (structures $B$ and $C$) in the non-equilibrium phase diagram (Fig.~\ref{non-equil-phase}) corresponding to $n_s=24$ and 26 do not constitute the minimum free energy structures. In the equilibrium phase diagram, they are both replaced by the structure $D$, a clathrin shell, which has $n_s=28$ and is called mini-coat.  The other clathrin shell, hexagonal barrel (structure $E$) obtained in the irreversible growth has $n_s=36$ with $D_{6h}$ dihedral symmetry. The equilibrium structure of the shell with the same $n_s=36$  has tennis ball symmetry, the structure $E^*$ in Figs.~\ref{equil-phase} and \ref{non-equil-phase}. 

The white area in the non-equilibrium phase diagram for $\gamma<2$ and between $1.21<R_0/b<1.3$ ($n_s=38-48$) corresponds to the region where many different types of shells without any specific symmetry are assembled. In contrast, there is no irregular structure in the equilibrium phase diagram, and we find the structures $I$ ($n_s=40$) and $J$ ($44$) with $D_2$ symmetry in that region. Furthermore, the regions corresponding to $G$ and $F$ structures cover a larger parameter space in the equilibrium phase diagram compared to the non-equilibrium one, revealing the presence of energy barriers preventing formation of some shells.

Furthermore, the irregular shells formed between the $G$ and $F$ structures at lower $\gamma$-values in the non-equilibrium phase diagram disappear and are replaced with the $F$ one. 
% Also structure $K$ with $n_s=76$ and tetrahedral symmetry appear between $G$ and $H$ (or $H^*$) that barely forms in non-equilibrium at the boundry between the hashed region and the white region in \ref{non-equil-phase}. (did not exist in the irreversible growth phase diagram).
The structure $K$ with $n_s=76$ and tetrahedral symmetry which forms between $G$ and $H$ (or $H^*$) in the equilibrium phase diagram, appears rarely in the non-equilibrium one. In fact it only assembles at the boundary between the hashed and the white regions (Fig. \ref{non-equil-phase}), despite the fact that $K$ structure is smaller than $H$ or $H^*$. Last but not least, the structures with icosahedral symmetry cover a wider region in the equilibrium phase diagram. The largest symmetric shell in Figs.~\ref{equil-phase} and \ref{non-equil-phase} is $n_s=80$. Note that in the absence of genome, an inner shell or scaffolding proteins, when we increase the spontaneous curvature at low $\gamma$ only irregular shapes form, and at large $\gamma$-values we obtain flat sheets or other structures with zero Gaussian curvature \cite{PhysRevE.74.031912,nguyen2006continuum,nature2016}. However, the focus of this paper is on the assembly of small symmetric shells, as illustrated in Figs.~\ref{equil-phase}.

% fig 9

% fig 10

\section{\label{sec:level5}Discussion and conclusion}

Despite the wide range of amino acid sequences and folding structures of coat proteins, many protein cages spontaneously self-assemble to form icosahedral or other symmetric structures. This reveals a ``universal'' behavior among most protein cages. In this paper, using the MC simulation and edge swapping method, we investigated the equilibrium structure of protein cages built from identical subunits.  We, in particular, choose triangular subunits as they are able to follow the general topological and geometrical constraints that all building blocks, no matter how complex or detailed, must obey. Since triangles form a hexagonal lattice in flat space and also assemble to create 12 disclinations for making closed polyhedrons, their behavior is similar to the building blocks of protein cages, {\it i.e.}, they belong to the same ``universality class'' as coat proteins.

Using triangular subunits, we studied the impact of the mechanical properties of building blocks on the symmetry and structure of small protein cages and constructed a phase diagram as a function of the spontaneous curvature of subunits and FvK number (the ratio of stretching to bending modulus), as shown in Fig.~\ref{equil-phase}.  The phase diagram is significantly occupied with icosahedral shells, $T=1$ and $3$, which are common among viruses and many other protein cages. As illustrated in the figure, at low $\gamma$, where subunits can deform easily from the equilateral triangle, various structures form as a function of spontaneous curvature. However, by increasing $\gamma$, subunits become more rigid and structures with lower symmetries disappear. For instance for low $\gamma=0.2$, the equilibrium structures are sensitive to the spontaneous radius of curvature and there is a smooth transition from one shell to the next one as illustrated in Fig.~\ref{N-R0-g0.2}. However by increasing to $\gamma=8$, the equilibrium structures become less sensitive to $R_0/b_0$ and only icosahedral structures with $T=1$ and 3 survive indicating the robustness of these two structures.

 Quite unexpectedly, we found that the equilibrium phase diagram, Fig.~\ref{equil-phase}, was very similar to the phase diagram obtained under non-equilibrium conditions, see Fig.~\ref{non-equil-phase}.   As explained in the previous section, the non-equilibrium simulations of Ref.~\cite{Wagner2015} were performed following the local minimum energy path but under the condition that once a pentamer or hexamer formed, it could no longer dissociate or move. Since the principles of detailed balance were violated in the irreversible growth simulations, we did not expect to observe such a high degree of similarity between the two phase diagrams, see Figs.~\ref{equil-phase} and \ref{non-equil-phase}.
 
These results could be explained to some extent with the recent work of Li {\it et al.} who employed the continuum elasticity theory and studied the assembly pathway of icosahedral shells. They found that as an elastic shell grows, there is a deep potential well attracting pentamers exactly at the locations that will become the vertices of an icosahedron when the shell is complete \cite{siyu2018}.  The extensive similarities between equilibrium and non-equilibrium phase diagrams in the current paper indicate that during the growth process, for the symmetric shells other than icosahedral ones there are also deep potential wells for the formation of disclinations. While we cannot solve the continuum elasticity equations in the presence of spontaneous curvature and explain the symmetry of shells observed in Figs.~\ref{structures} and \ref{structures2} but based on our results, it is obvious that there is a delta-type potential for the formation of disclinations at specific locations during the growth process, which leads to the assembly of different types of symmetric shells depending on the physical properties of protein subunits.

It is important to note that we often found one single global minimum energy structure in the phase diagrams presented in Figs.~\ref{structures} and \ref{structures2} for a given $\gamma$ and $R_0/b_0$. However, in many biological systems, sometimes a few different types of protein cages co-exist in the same solution \cite{Hsiang-Ku,Kusters2015}. For instance self-assembly studies of dimeric Hepatitis B Virus capsid protein mutant Cp1492 shows that empty $T=3$ and $T=4$ structures form in a ratio of about 95:5 at medium to high salt concentration and close to neutral pH \cite{moerman2016kinetics, wingfield1995hepatitis}.  While the focus of our work is to find the optimal structure of protein cages as a function of mechanical properties of its building blocks, the polymorphism observed in several self-assembly studies can be explained through a careful examination of plots of energy per subunit versus number of subunits in Fig.~\ref{Energy vs Nv}.  If the difference between the free energy per subunits in two different structures is small compared to the thermal energy $k_BT$, one expects to observe both structures, with relative populations given by the corresponding Boltzmann factor $\exp(\Delta \epsilon/k_BT)$, with $\Delta \epsilon$ the difference between the free energy per subunit in the two shells.  

Lastly we emphasize that even though in this work we did not explicitly study the impact of salt and pH on the structure of protein shells, the solution environment such as salt and pH can modify the number of charges on the protein subunits, which in turn can change the stiffness and the spontaneous dihedral angle of building blocks.  While the results obtained in this paper can explain why various protein shells with different symmetry appear in nature, at this point there is not enough experimental data to allow us to connect our variables $R_0/b_0$ and $\gamma$ (the ratio of stretching to bending modulus) to the experimental conditions such as pH and salt.  
%Depending on the condition of solution and stiffness of proteins, the free energy per subunit might be such that a few different structures co-exist in the same solution as observed in the self-assembly studies of clathrin shells \cite{fotin2004molecular}. 

 To examine several concepts presented in this article, it would be interesting to carry out a set of systematic experiments as a function of pH and salt concentration with various mutated proteins, which in consequence have different mechanical properties.  One then can construct an experimental phase diagram similar to the one shown in Fig.~\ref{equil-phase}.  A quantitative comparison between experiments and the theory will result in a better understanding of the protein-protein interaction and the parameters that contribute to the formation of various protein cages with extensive application in various area of material science, gene delivery and medicine. 

\section{\label{sec:level6}Acknowledgement}
This work was supported by the National Science Foundation through Grant No. DMR-1719550.

% \subsubsection{Wide text (A level-3 head)}
% The \texttt{widetext} environment will make the text the width of the
% full page, as on page~\pageref{eq:wideeq}. (Note the use the
% \verb+\pageref{#1}+ command to refer to the page number.) 
% \paragraph{Note (Fourth-level head is run in)}
% The width-changing commands only take effect in two-column formatting. 
% There is no effect if text is in a single column.

%\subsection{\label{sec:citeref}Citations and References}

% A citation in text uses the command \verb+\cite{#1}+ or
% \verb+\onlinecite{#1}+ and refers to an entry in the bibliography. 
% An entry in the bibliography is a reference to another document.

\bibliography{bibfile}

\begin{thebibliography}{42}%
\makeatletter
\providecommand \@ifxundefined [1]{%
 \@ifx{#1\undefined}
}%
\providecommand \@ifnum [1]{%
 \ifnum #1\expandafter \@firstoftwo
 \else \expandafter \@secondoftwo
 \fi
}%
\providecommand \@ifx [1]{%
 \ifx #1\expandafter \@firstoftwo
 \else \expandafter \@secondoftwo
 \fi
}%
\providecommand \natexlab [1]{#1}%
\providecommand \enquote  [1]{``#1''}%
\providecommand \bibnamefont  [1]{#1}%
\providecommand \bibfnamefont [1]{#1}%
\providecommand \citenamefont [1]{#1}%
\providecommand \href@noop [0]{\@secondoftwo}%
\providecommand \href [0]{\begingroup \@sanitize@url \@href}%
\providecommand \@href[1]{\@@startlink{#1}\@@href}%
\providecommand \@@href[1]{\endgroup#1\@@endlink}%
\providecommand \@sanitize@url [0]{\catcode `\\12\catcode `\$12\catcode
  `\&12\catcode `\#12\catcode `\^12\catcode `\_12\catcode `\%12\relax}%
\providecommand \@@startlink[1]{}%
\providecommand \@@endlink[0]{}%
\providecommand \url  [0]{\begingroup\@sanitize@url \@url }%
\providecommand \@url [1]{\endgroup\@href {#1}{\urlprefix }}%
\providecommand \urlprefix  [0]{URL }%
\providecommand \Eprint [0]{\href }%
\providecommand \doibase [0]{http://dx.doi.org/}%
\providecommand \selectlanguage [0]{\@gobble}%
\providecommand \bibinfo  [0]{\@secondoftwo}%
\providecommand \bibfield  [0]{\@secondoftwo}%
\providecommand \translation [1]{[#1]}%
\providecommand \BibitemOpen [0]{}%
\providecommand \bibitemStop [0]{}%
\providecommand \bibitemNoStop [0]{.\EOS\space}%
\providecommand \EOS [0]{\spacefactor3000\relax}%
\providecommand \BibitemShut  [1]{\csname bibitem#1\endcsname}%
\let\auto@bib@innerbib\@empty
%</preamble>
\bibitem [{\citenamefont {Kroto}\ \emph {et~al.}(1985)\citenamefont {Kroto},
  \citenamefont {Heath}, \citenamefont {O'Brien}, \citenamefont {Curl},\ and\
  \citenamefont {Smalley}}]{kroto1985c60}%
  \BibitemOpen
  \bibfield  {author} {\bibinfo {author} {\bibfnamefont {H.~W.}\ \bibnamefont
  {Kroto}}, \bibinfo {author} {\bibfnamefont {J.~R.}\ \bibnamefont {Heath}},
  \bibinfo {author} {\bibfnamefont {S.~C.}\ \bibnamefont {O'Brien}}, \bibinfo
  {author} {\bibfnamefont {R.~F.}\ \bibnamefont {Curl}}, \ and\ \bibinfo
  {author} {\bibfnamefont {R.~E.}\ \bibnamefont {Smalley}},\ }\href@noop {}
  {\bibfield  {journal} {\bibinfo  {journal} {Nature}\ }\textbf {\bibinfo
  {volume} {318}},\ \bibinfo {pages} {162} (\bibinfo {year}
  {1985})}\BibitemShut {NoStop}%
\bibitem [{\citenamefont {Baughman}\ \emph {et~al.}(1993)\citenamefont
  {Baughman}, \citenamefont {Galv{\~a}o}, \citenamefont {Cui}, \citenamefont
  {Wang},\ and\ \citenamefont {Tom{\'a}nek}}]{baughman1993fullereneynes}%
  \BibitemOpen
  \bibfield  {author} {\bibinfo {author} {\bibfnamefont {R.~H.}\ \bibnamefont
  {Baughman}}, \bibinfo {author} {\bibfnamefont {D.~S.}\ \bibnamefont
  {Galv{\~a}o}}, \bibinfo {author} {\bibfnamefont {C.}~\bibnamefont {Cui}},
  \bibinfo {author} {\bibfnamefont {Y.}~\bibnamefont {Wang}}, \ and\ \bibinfo
  {author} {\bibfnamefont {D.}~\bibnamefont {Tom{\'a}nek}},\ }\href@noop {}
  {\bibfield  {journal} {\bibinfo  {journal} {Chemical physics letters}\
  }\textbf {\bibinfo {volume} {204}},\ \bibinfo {pages} {8} (\bibinfo {year}
  {1993})}\BibitemShut {NoStop}%
\bibitem [{\citenamefont {Aumiller}\ \emph {et~al.}(2018)\citenamefont
  {Aumiller}, \citenamefont {Uchida},\ and\ \citenamefont
  {Douglas}}]{aumiller2018protein}%
  \BibitemOpen
  \bibfield  {author} {\bibinfo {author} {\bibfnamefont {W.~M.}\ \bibnamefont
  {Aumiller}}, \bibinfo {author} {\bibfnamefont {M.}~\bibnamefont {Uchida}}, \
  and\ \bibinfo {author} {\bibfnamefont {T.}~\bibnamefont {Douglas}},\
  }\href@noop {} {\bibfield  {journal} {\bibinfo  {journal} {Chemical Society
  Reviews}\ }\textbf {\bibinfo {volume} {47}},\ \bibinfo {pages} {3433}
  (\bibinfo {year} {2018})}\BibitemShut {NoStop}%
\bibitem [{\citenamefont {Fontana}\ \emph {et~al.}(2014)\citenamefont
  {Fontana}, \citenamefont {Nemecek}, \citenamefont {McHugh}, \citenamefont
  {Aksyuk}, \citenamefont {Cheng}, \citenamefont {Winkler}, \citenamefont
  {Heymann}, \citenamefont {Hoiczyk},\ and\ \citenamefont
  {Steven}}]{fontana2014phage}%
  \BibitemOpen
  \bibfield  {author} {\bibinfo {author} {\bibfnamefont {J.}~\bibnamefont
  {Fontana}}, \bibinfo {author} {\bibfnamefont {D.}~\bibnamefont {Nemecek}},
  \bibinfo {author} {\bibfnamefont {C.~A.}\ \bibnamefont {McHugh}}, \bibinfo
  {author} {\bibfnamefont {A.~A.}\ \bibnamefont {Aksyuk}}, \bibinfo {author}
  {\bibfnamefont {N.}~\bibnamefont {Cheng}}, \bibinfo {author} {\bibfnamefont
  {D.~C.}\ \bibnamefont {Winkler}}, \bibinfo {author} {\bibfnamefont {J.~B.}\
  \bibnamefont {Heymann}}, \bibinfo {author} {\bibfnamefont {E.}~\bibnamefont
  {Hoiczyk}}, \ and\ \bibinfo {author} {\bibfnamefont {A.~C.}\ \bibnamefont
  {Steven}},\ }\href@noop {} {\bibfield  {journal} {\bibinfo  {journal}
  {Microscopy and Microanalysis}\ }\textbf {\bibinfo {volume} {20}},\ \bibinfo
  {pages} {1244} (\bibinfo {year} {2014})}\BibitemShut {NoStop}%
\bibitem [{\citenamefont {Ma}\ \emph {et~al.}(2018)\citenamefont {Ma},
  \citenamefont {Gong}, \citenamefont {Aubert}, \citenamefont {Turker},
  \citenamefont {Kao}, \citenamefont {Doerschuk},\ and\ \citenamefont
  {Wiesner}}]{ma2018self}%
  \BibitemOpen
  \bibfield  {author} {\bibinfo {author} {\bibfnamefont {K.}~\bibnamefont
  {Ma}}, \bibinfo {author} {\bibfnamefont {Y.}~\bibnamefont {Gong}}, \bibinfo
  {author} {\bibfnamefont {T.}~\bibnamefont {Aubert}}, \bibinfo {author}
  {\bibfnamefont {M.~Z.}\ \bibnamefont {Turker}}, \bibinfo {author}
  {\bibfnamefont {T.}~\bibnamefont {Kao}}, \bibinfo {author} {\bibfnamefont
  {P.~C.}\ \bibnamefont {Doerschuk}}, \ and\ \bibinfo {author} {\bibfnamefont
  {U.}~\bibnamefont {Wiesner}},\ }\href@noop {} {\bibfield  {journal} {\bibinfo
   {journal} {Nature}\ ,\ \bibinfo {pages} {1}} (\bibinfo {year}
  {2018})}\BibitemShut {NoStop}%
\bibitem [{\citenamefont {Perlmutter}\ \emph {et~al.}(2013)\citenamefont
  {Perlmutter}, \citenamefont {Qiao},\ and\ \citenamefont {Hagan}}]{elife}%
  \BibitemOpen
  \bibfield  {author} {\bibinfo {author} {\bibfnamefont {J.~D.}\ \bibnamefont
  {Perlmutter}}, \bibinfo {author} {\bibfnamefont {C.}~\bibnamefont {Qiao}}, \
  and\ \bibinfo {author} {\bibfnamefont {M.~F.}\ \bibnamefont {Hagan}},\ }\href
  {\doibase 10.7554/eLife.00632} {\bibfield  {journal} {\bibinfo  {journal}
  {Elife}\ }\textbf {\bibinfo {volume} {2}} (\bibinfo {year} {2013}),\
  10.7554/eLife.00632}\BibitemShut {NoStop}%
\bibitem [{\citenamefont {Erdemci-Tandogan}\ \emph {et~al.}(2016)\citenamefont
  {Erdemci-Tandogan}, \citenamefont {Wagner}, \citenamefont {van~der Schoot},
  \citenamefont {Podgornik},\ and\ \citenamefont {Zandi}}]{Gonca2016}%
  \BibitemOpen
  \bibfield  {author} {\bibinfo {author} {\bibfnamefont {G.}~\bibnamefont
  {Erdemci-Tandogan}}, \bibinfo {author} {\bibfnamefont {J.}~\bibnamefont
  {Wagner}}, \bibinfo {author} {\bibfnamefont {P.}~\bibnamefont {van~der
  Schoot}}, \bibinfo {author} {\bibfnamefont {R.}~\bibnamefont {Podgornik}}, \
  and\ \bibinfo {author} {\bibfnamefont {R.}~\bibnamefont {Zandi}},\
  }\href@noop {} {\bibfield  {journal} {\bibinfo  {journal} {Physical Review
  E}\ }\textbf {\bibinfo {volume} {94}},\ \bibinfo {pages} {022408} (\bibinfo
  {year} {2016})}\BibitemShut {NoStop}%
\bibitem [{\citenamefont {Heddle}(2008)}]{heddle2008protein}%
  \BibitemOpen
  \bibfield  {author} {\bibinfo {author} {\bibfnamefont {J.~G.}\ \bibnamefont
  {Heddle}},\ }\href@noop {} {\bibfield  {journal} {\bibinfo  {journal}
  {Nanotechnology, science and applications}\ }\textbf {\bibinfo {volume}
  {1}},\ \bibinfo {pages} {67} (\bibinfo {year} {2008})}\BibitemShut {NoStop}%
\bibitem [{\citenamefont {Garmann}\ \emph {et~al.}(2015)\citenamefont
  {Garmann}, \citenamefont {Comas-Garcia}, \citenamefont {Knobler},\ and\
  \citenamefont {Gelbart}}]{Garmann2015}%
  \BibitemOpen
  \bibfield  {author} {\bibinfo {author} {\bibfnamefont {R.~F.}\ \bibnamefont
  {Garmann}}, \bibinfo {author} {\bibfnamefont {M.}~\bibnamefont
  {Comas-Garcia}}, \bibinfo {author} {\bibfnamefont {C.~M.}\ \bibnamefont
  {Knobler}}, \ and\ \bibinfo {author} {\bibfnamefont {W.~M.}\ \bibnamefont
  {Gelbart}},\ }\href@noop {} {\bibfield  {journal} {\bibinfo  {journal}
  {Accounts of chemical research}\ }\textbf {\bibinfo {volume} {49}},\ \bibinfo
  {pages} {48} (\bibinfo {year} {2015})}\BibitemShut {NoStop}%
\bibitem [{\citenamefont {Chevreuil}\ \emph {et~al.}(2018)\citenamefont
  {Chevreuil}, \citenamefont {Law-Hine}, \citenamefont {Chen}, \citenamefont
  {Bressanelli}, \citenamefont {Combet}, \citenamefont {Constantin},
  \citenamefont {Degrouard}, \citenamefont {M{\"o}ller}, \citenamefont
  {Zeghal},\ and\ \citenamefont {Tresset}}]{chevreuil2018nonequilibrium}%
  \BibitemOpen
  \bibfield  {author} {\bibinfo {author} {\bibfnamefont {M.}~\bibnamefont
  {Chevreuil}}, \bibinfo {author} {\bibfnamefont {D.}~\bibnamefont {Law-Hine}},
  \bibinfo {author} {\bibfnamefont {J.}~\bibnamefont {Chen}}, \bibinfo {author}
  {\bibfnamefont {S.}~\bibnamefont {Bressanelli}}, \bibinfo {author}
  {\bibfnamefont {S.}~\bibnamefont {Combet}}, \bibinfo {author} {\bibfnamefont
  {D.}~\bibnamefont {Constantin}}, \bibinfo {author} {\bibfnamefont
  {J.}~\bibnamefont {Degrouard}}, \bibinfo {author} {\bibfnamefont
  {J.}~\bibnamefont {M{\"o}ller}}, \bibinfo {author} {\bibfnamefont
  {M.}~\bibnamefont {Zeghal}}, \ and\ \bibinfo {author} {\bibfnamefont
  {G.}~\bibnamefont {Tresset}},\ }\href@noop {} {\bibfield  {journal} {\bibinfo
   {journal} {Nature communications}\ }\textbf {\bibinfo {volume} {9}},\
  \bibinfo {pages} {3071} (\bibinfo {year} {2018})}\BibitemShut {NoStop}%
\bibitem [{\citenamefont {Crick}\ and\ \citenamefont
  {Watson}(1956)}]{crick1956structure}%
  \BibitemOpen
  \bibfield  {author} {\bibinfo {author} {\bibfnamefont {F.~H.}\ \bibnamefont
  {Crick}}\ and\ \bibinfo {author} {\bibfnamefont {J.~D.}\ \bibnamefont
  {Watson}},\ }\href@noop {} {\bibfield  {journal} {\bibinfo  {journal}
  {Nature}\ }\textbf {\bibinfo {volume} {177}},\ \bibinfo {pages} {473}
  (\bibinfo {year} {1956})}\BibitemShut {NoStop}%
\bibitem [{\citenamefont {Weiss}(2005)}]{weiss2005armor}%
  \BibitemOpen
  \bibfield  {author} {\bibinfo {author} {\bibfnamefont {P.}~\bibnamefont
  {Weiss}},\ }\href@noop {} {\bibfield  {journal} {\bibinfo  {journal} {Science
  News}\ }\textbf {\bibinfo {volume} {168}},\ \bibinfo {pages} {152} (\bibinfo
  {year} {2005})}\BibitemShut {NoStop}%
\bibitem [{\citenamefont {Johnson}\ and\ \citenamefont
  {Speir}(1997)}]{johnson1997quasi}%
  \BibitemOpen
  \bibfield  {author} {\bibinfo {author} {\bibfnamefont {J.~E.}\ \bibnamefont
  {Johnson}}\ and\ \bibinfo {author} {\bibfnamefont {J.~A.}\ \bibnamefont
  {Speir}},\ }\href@noop {} {\bibfield  {journal} {\bibinfo  {journal} {Journal
  of molecular biology}\ }\textbf {\bibinfo {volume} {269}},\ \bibinfo {pages}
  {665} (\bibinfo {year} {1997})}\BibitemShut {NoStop}%
\bibitem [{\citenamefont {Caspar}\ and\ \citenamefont
  {Klug}(1962)}]{CASPAR1962}%
  \BibitemOpen
  \bibfield  {author} {\bibinfo {author} {\bibfnamefont {D.~L.}\ \bibnamefont
  {Caspar}}\ and\ \bibinfo {author} {\bibfnamefont {A.}~\bibnamefont {Klug}},\
  }\href {\doibase doi:10.1101/SQB.1962.027.001.005} {\bibfield  {journal}
  {\bibinfo  {journal} {Cold Spring Harbor Symp. Quant. Biol.}\ }\textbf
  {\bibinfo {volume} {27}},\ \bibinfo {pages} {1} (\bibinfo {year}
  {1962})}\BibitemShut {NoStop}%
\bibitem [{\citenamefont {Cheng}\ \emph {et~al.}(2007)\citenamefont {Cheng},
  \citenamefont {Boll}, \citenamefont {Kirchhausen}, \citenamefont {Harrison},\
  and\ \citenamefont {Walz}}]{cheng2007cryo}%
  \BibitemOpen
  \bibfield  {author} {\bibinfo {author} {\bibfnamefont {Y.}~\bibnamefont
  {Cheng}}, \bibinfo {author} {\bibfnamefont {W.}~\bibnamefont {Boll}},
  \bibinfo {author} {\bibfnamefont {T.}~\bibnamefont {Kirchhausen}}, \bibinfo
  {author} {\bibfnamefont {S.~C.}\ \bibnamefont {Harrison}}, \ and\ \bibinfo
  {author} {\bibfnamefont {T.}~\bibnamefont {Walz}},\ }\href@noop {} {\bibfield
   {journal} {\bibinfo  {journal} {Journal of molecular biology}\ }\textbf
  {\bibinfo {volume} {365}},\ \bibinfo {pages} {892} (\bibinfo {year}
  {2007})}\BibitemShut {NoStop}%
\bibitem [{\citenamefont {Crowther}\ \emph {et~al.}(1976)\citenamefont
  {Crowther}, \citenamefont {Pinch},\ and\ \citenamefont
  {Pearse}}]{crowther1976structure}%
  \BibitemOpen
  \bibfield  {author} {\bibinfo {author} {\bibfnamefont {R.}~\bibnamefont
  {Crowther}}, \bibinfo {author} {\bibfnamefont {J.}~\bibnamefont {Pinch}}, \
  and\ \bibinfo {author} {\bibfnamefont {B.}~\bibnamefont {Pearse}},\
  }\href@noop {} {\bibfield  {journal} {\bibinfo  {journal} {Journal of
  molecular biology}\ }\textbf {\bibinfo {volume} {103}},\ \bibinfo {pages}
  {785} (\bibinfo {year} {1976})}\BibitemShut {NoStop}%
\bibitem [{\citenamefont {Min}\ \emph {et~al.}(2014)\citenamefont {Min},
  \citenamefont {Kim}, \citenamefont {Lee},\ and\ \citenamefont
  {Kang}}]{min2014lumazine}%
  \BibitemOpen
  \bibfield  {author} {\bibinfo {author} {\bibfnamefont {J.}~\bibnamefont
  {Min}}, \bibinfo {author} {\bibfnamefont {S.}~\bibnamefont {Kim}}, \bibinfo
  {author} {\bibfnamefont {J.}~\bibnamefont {Lee}}, \ and\ \bibinfo {author}
  {\bibfnamefont {S.}~\bibnamefont {Kang}},\ }\href@noop {} {\bibfield
  {journal} {\bibinfo  {journal} {RSC Advances}\ }\textbf {\bibinfo {volume}
  {4}},\ \bibinfo {pages} {48596} (\bibinfo {year} {2014})}\BibitemShut
  {NoStop}%
\bibitem [{\citenamefont {Fotin}\ \emph {et~al.}(2004)\citenamefont {Fotin},
  \citenamefont {Cheng}, \citenamefont {Sliz}, \citenamefont {Grigorieff},
  \citenamefont {Harrison}, \citenamefont {Kirchhausen},\ and\ \citenamefont
  {Walz}}]{fotin2004molecular}%
  \BibitemOpen
  \bibfield  {author} {\bibinfo {author} {\bibfnamefont {A.}~\bibnamefont
  {Fotin}}, \bibinfo {author} {\bibfnamefont {Y.}~\bibnamefont {Cheng}},
  \bibinfo {author} {\bibfnamefont {P.}~\bibnamefont {Sliz}}, \bibinfo {author}
  {\bibfnamefont {N.}~\bibnamefont {Grigorieff}}, \bibinfo {author}
  {\bibfnamefont {S.~C.}\ \bibnamefont {Harrison}}, \bibinfo {author}
  {\bibfnamefont {T.}~\bibnamefont {Kirchhausen}}, \ and\ \bibinfo {author}
  {\bibfnamefont {T.}~\bibnamefont {Walz}},\ }\href@noop {} {\bibfield
  {journal} {\bibinfo  {journal} {nature}\ }\textbf {\bibinfo {volume} {432}},\
  \bibinfo {pages} {573} (\bibinfo {year} {2004})}\BibitemShut {NoStop}%
\bibitem [{\citenamefont {McHugh}\ \emph {et~al.}(2014)\citenamefont {McHugh},
  \citenamefont {Fontana}, \citenamefont {Nemecek}, \citenamefont {Cheng},
  \citenamefont {Aksyuk}, \citenamefont {Heymann}, \citenamefont {Winkler},
  \citenamefont {Lam}, \citenamefont {Wall}, \citenamefont {Steven},\ and\
  \citenamefont {Hoiczyk}}]{mchugh2014virus}%
  \BibitemOpen
  \bibfield  {author} {\bibinfo {author} {\bibfnamefont {C.~A.}\ \bibnamefont
  {McHugh}}, \bibinfo {author} {\bibfnamefont {J.}~\bibnamefont {Fontana}},
  \bibinfo {author} {\bibfnamefont {D.}~\bibnamefont {Nemecek}}, \bibinfo
  {author} {\bibfnamefont {N.}~\bibnamefont {Cheng}}, \bibinfo {author}
  {\bibfnamefont {A.~A.}\ \bibnamefont {Aksyuk}}, \bibinfo {author}
  {\bibfnamefont {J.~B.}\ \bibnamefont {Heymann}}, \bibinfo {author}
  {\bibfnamefont {D.~C.}\ \bibnamefont {Winkler}}, \bibinfo {author}
  {\bibfnamefont {A.~S.}\ \bibnamefont {Lam}}, \bibinfo {author} {\bibfnamefont
  {J.~S.}\ \bibnamefont {Wall}}, \bibinfo {author} {\bibfnamefont {A.~C.}\
  \bibnamefont {Steven}}, \ and\ \bibinfo {author} {\bibfnamefont
  {E.}~\bibnamefont {Hoiczyk}},\ }\href@noop {} {\bibfield  {journal} {\bibinfo
   {journal} {The EMBO journal}\ }\textbf {\bibinfo {volume} {33}},\ \bibinfo
  {pages} {1896} (\bibinfo {year} {2014})}\BibitemShut {NoStop}%
\bibitem [{\citenamefont {Hagan}\ and\ \citenamefont
  {Zandi}(2016)}]{Zandi2016}%
  \BibitemOpen
  \bibfield  {author} {\bibinfo {author} {\bibfnamefont {M.~F.}\ \bibnamefont
  {Hagan}}\ and\ \bibinfo {author} {\bibfnamefont {R.}~\bibnamefont {Zandi}},\
  }\href {\doibase 10.1016/j.coviro.2016.02.012} {\bibfield  {journal}
  {\bibinfo  {journal} {Curr. Opin. Virol.}\ }\textbf {\bibinfo {volume}
  {18}},\ \bibinfo {pages} {36} (\bibinfo {year} {2016})}\BibitemShut {NoStop}%
\bibitem [{\citenamefont {Chen}\ \emph
  {et~al.}(2007{\natexlab{a}})\citenamefont {Chen}, \citenamefont {Zhang},\
  and\ \citenamefont {Glotzer}}]{doi:10.1021/la063755d}%
  \BibitemOpen
  \bibfield  {author} {\bibinfo {author} {\bibfnamefont {T.}~\bibnamefont
  {Chen}}, \bibinfo {author} {\bibfnamefont {Z.}~\bibnamefont {Zhang}}, \ and\
  \bibinfo {author} {\bibfnamefont {S.~C.}\ \bibnamefont {Glotzer}},\ }\href
  {\doibase 10.1021/la063755d} {\bibfield  {journal} {\bibinfo  {journal}
  {Langmuir}\ }\textbf {\bibinfo {volume} {23}},\ \bibinfo {pages} {6598}
  (\bibinfo {year} {2007}{\natexlab{a}})},\ \bibinfo {note} {pMID: 17489618},\
  \Eprint {http://arxiv.org/abs/https://doi.org/10.1021/la063755d}
  {https://doi.org/10.1021/la063755d} \BibitemShut {NoStop}%
\bibitem [{\citenamefont {Chen}\ \emph
  {et~al.}(2007{\natexlab{b}})\citenamefont {Chen}, \citenamefont {Zhang},\
  and\ \citenamefont {Glotzer}}]{Chen:2007b}%
  \BibitemOpen
  \bibfield  {author} {\bibinfo {author} {\bibfnamefont {T.}~\bibnamefont
  {Chen}}, \bibinfo {author} {\bibfnamefont {Z.}~\bibnamefont {Zhang}}, \ and\
  \bibinfo {author} {\bibfnamefont {S.~C.}\ \bibnamefont {Glotzer}},\
  }\href@noop {} {\bibfield  {journal} {\bibinfo  {journal} {Proceedings of the
  National Academy of Sciences}\ }\textbf {\bibinfo {volume} {104}},\ \bibinfo
  {pages} {717} (\bibinfo {year} {2007}{\natexlab{b}})}\BibitemShut {NoStop}%
\bibitem [{\citenamefont {Zandi}\ \emph {et~al.}(2004)\citenamefont {Zandi},
  \citenamefont {Reguera}, \citenamefont {Bruinsma}, \citenamefont {Gelbart},\
  and\ \citenamefont {Rudnick}}]{zandi2004}%
  \BibitemOpen
  \bibfield  {author} {\bibinfo {author} {\bibfnamefont {R.}~\bibnamefont
  {Zandi}}, \bibinfo {author} {\bibfnamefont {D.}~\bibnamefont {Reguera}},
  \bibinfo {author} {\bibfnamefont {R.~F.}\ \bibnamefont {Bruinsma}}, \bibinfo
  {author} {\bibfnamefont {W.~M.}\ \bibnamefont {Gelbart}}, \ and\ \bibinfo
  {author} {\bibfnamefont {J.}~\bibnamefont {Rudnick}},\ }\href@noop {}
  {\bibfield  {journal} {\bibinfo  {journal} {Proceedings of the National
  Academy of Sciences}\ }\textbf {\bibinfo {volume} {101}},\ \bibinfo {pages}
  {15556} (\bibinfo {year} {2004})}\BibitemShut {NoStop}%
\bibitem [{\citenamefont {Paquay}\ \emph {et~al.}(2016)\citenamefont {Paquay},
  \citenamefont {Kusumaatmaja}, \citenamefont {Wales}, \citenamefont {Zandi},\
  and\ \citenamefont {van~der Schoot}}]{Stefan}%
  \BibitemOpen
  \bibfield  {author} {\bibinfo {author} {\bibfnamefont {S.}~\bibnamefont
  {Paquay}}, \bibinfo {author} {\bibfnamefont {H.}~\bibnamefont
  {Kusumaatmaja}}, \bibinfo {author} {\bibfnamefont {D.~J.}\ \bibnamefont
  {Wales}}, \bibinfo {author} {\bibfnamefont {R.}~\bibnamefont {Zandi}}, \ and\
  \bibinfo {author} {\bibfnamefont {P.}~\bibnamefont {van~der Schoot}},\ }\href
  {\doibase 10.1039/C6SM00489J} {\bibfield  {journal} {\bibinfo  {journal}
  {Soft Matter}\ }\textbf {\bibinfo {volume} {12}},\ \bibinfo {pages} {5708}
  (\bibinfo {year} {2016})}\BibitemShut {NoStop}%
\bibitem [{\citenamefont {Wagner}\ and\ \citenamefont
  {Zandi}(2015)}]{Wagner2015}%
  \BibitemOpen
  \bibfield  {author} {\bibinfo {author} {\bibfnamefont {J.}~\bibnamefont
  {Wagner}}\ and\ \bibinfo {author} {\bibfnamefont {R.}~\bibnamefont {Zandi}},\
  }\href@noop {} {\bibfield  {journal} {\bibinfo  {journal} {Biophysical
  journal}\ }\textbf {\bibinfo {volume} {109}},\ \bibinfo {pages} {956}
  (\bibinfo {year} {2015})}\BibitemShut {NoStop}%
\bibitem [{\citenamefont {Kohyama}\ \emph {et~al.}(2003)\citenamefont
  {Kohyama}, \citenamefont {Kroll},\ and\ \citenamefont
  {Gompper}}]{kohyama2003budding}%
  \BibitemOpen
  \bibfield  {author} {\bibinfo {author} {\bibfnamefont {T.}~\bibnamefont
  {Kohyama}}, \bibinfo {author} {\bibfnamefont {D.}~\bibnamefont {Kroll}}, \
  and\ \bibinfo {author} {\bibfnamefont {G.}~\bibnamefont {Gompper}},\
  }\href@noop {} {\bibfield  {journal} {\bibinfo  {journal} {Physical Review
  E}\ }\textbf {\bibinfo {volume} {68}},\ \bibinfo {pages} {061905} (\bibinfo
  {year} {2003})}\BibitemShut {NoStop}%
\bibitem [{\citenamefont {Rotskoff}\ and\ \citenamefont
  {Geissler}(2018)}]{rotskoff2018robust}%
  \BibitemOpen
  \bibfield  {author} {\bibinfo {author} {\bibfnamefont {G.~M.}\ \bibnamefont
  {Rotskoff}}\ and\ \bibinfo {author} {\bibfnamefont {P.~L.}\ \bibnamefont
  {Geissler}},\ }\href@noop {} {\bibfield  {journal} {\bibinfo  {journal}
  {Proceedings of the National Academy of Sciences}\ }\textbf {\bibinfo
  {volume} {115}},\ \bibinfo {pages} {6341} (\bibinfo {year}
  {2018})}\BibitemShut {NoStop}%
\bibitem [{\citenamefont {Fejer}\ \emph {et~al.}(2010)\citenamefont {Fejer},
  \citenamefont {Chakrabarti},\ and\ \citenamefont {Wales}}]{Fejer:10}%
  \BibitemOpen
  \bibfield  {author} {\bibinfo {author} {\bibfnamefont {S.}~\bibnamefont
  {Fejer}}, \bibinfo {author} {\bibfnamefont {D.}~\bibnamefont {Chakrabarti}},
  \ and\ \bibinfo {author} {\bibfnamefont {D.}~\bibnamefont {Wales}},\
  }\href@noop {} {\bibfield  {journal} {\bibinfo  {journal} {Nano Lett}\
  }\textbf {\bibinfo {volume} {4}},\ \bibinfo {pages} {219} (\bibinfo {year}
  {2010})}\BibitemShut {NoStop}%
\bibitem [{\citenamefont {Sivanandam}\ \emph {et~al.}(2016)\citenamefont
  {Sivanandam}, \citenamefont {Mathews}, \citenamefont {Garmann}, \citenamefont
  {Erdemci-Tandogan}, \citenamefont {Zandi},\ and\ \citenamefont
  {Rao}}]{Venky2016}%
  \BibitemOpen
  \bibfield  {author} {\bibinfo {author} {\bibfnamefont {V.}~\bibnamefont
  {Sivanandam}}, \bibinfo {author} {\bibfnamefont {D.}~\bibnamefont {Mathews}},
  \bibinfo {author} {\bibfnamefont {R.}~\bibnamefont {Garmann}}, \bibinfo
  {author} {\bibfnamefont {G.}~\bibnamefont {Erdemci-Tandogan}}, \bibinfo
  {author} {\bibfnamefont {R.}~\bibnamefont {Zandi}}, \ and\ \bibinfo {author}
  {\bibfnamefont {A.~L.~N.}\ \bibnamefont {Rao}},\ }\href
  {http://dx.doi.org/10.1038/srep26328 http://10.1038/srep26328} {\bibfield
  {journal} {\bibinfo  {journal} {Scientific Reports}\ }\textbf {\bibinfo
  {volume} {6}},\ \bibinfo {pages} {26328} (\bibinfo {year}
  {2016})}\BibitemShut {NoStop}%
\bibitem [{\citenamefont {Hicks}\ and\ \citenamefont
  {Henley}(2006)}]{PhysRevE.74.031912}%
  \BibitemOpen
  \bibfield  {author} {\bibinfo {author} {\bibfnamefont {S.~D.}\ \bibnamefont
  {Hicks}}\ and\ \bibinfo {author} {\bibfnamefont {C.~L.}\ \bibnamefont
  {Henley}},\ }\href {\doibase 10.1103/PhysRevE.74.031912} {\bibfield
  {journal} {\bibinfo  {journal} {Phys. Rev. E}\ }\textbf {\bibinfo {volume}
  {74}},\ \bibinfo {pages} {031912} (\bibinfo {year} {2006})}\BibitemShut
  {NoStop}%
\bibitem [{\citenamefont {Li}\ and\ \citenamefont {Scheraga}(1987)}]{Li6611}%
  \BibitemOpen
  \bibfield  {author} {\bibinfo {author} {\bibfnamefont {Z.}~\bibnamefont
  {Li}}\ and\ \bibinfo {author} {\bibfnamefont {H.~A.}\ \bibnamefont
  {Scheraga}},\ }\href {\doibase 10.1073/pnas.84.19.6611} {\bibfield  {journal}
  {\bibinfo  {journal} {Proceedings of the National Academy of Sciences}\
  }\textbf {\bibinfo {volume} {84}},\ \bibinfo {pages} {6611} (\bibinfo {year}
  {1987})},\ \Eprint
  {http://arxiv.org/abs/http://www.pnas.org/content/84/19/6611.full.pdf}
  {http://www.pnas.org/content/84/19/6611.full.pdf} \BibitemShut {NoStop}%
\bibitem [{\citenamefont {Nocedal}\ and\ \citenamefont
  {Wright}(2006)}]{NoceWrig06}%
  \BibitemOpen
  \bibfield  {author} {\bibinfo {author} {\bibfnamefont {J.}~\bibnamefont
  {Nocedal}}\ and\ \bibinfo {author} {\bibfnamefont {S.~J.}\ \bibnamefont
  {Wright}},\ }\href@noop {} {\emph {\bibinfo {title} {Numerical
  Optimization}}},\ \bibinfo {edition} {2nd}\ ed.\ (\bibinfo  {publisher}
  {Springer},\ \bibinfo {address} {New York, NY, USA},\ \bibinfo {year}
  {2006})\BibitemShut {NoStop}%
\bibitem [{\citenamefont {Dokland}(1999)}]{dokland1999scaffolding}%
  \BibitemOpen
  \bibfield  {author} {\bibinfo {author} {\bibfnamefont {T.}~\bibnamefont
  {Dokland}},\ }\href@noop {} {\bibfield  {journal} {\bibinfo  {journal}
  {Cellular and Molecular Life Sciences CMLS}\ }\textbf {\bibinfo {volume}
  {56}},\ \bibinfo {pages} {580} (\bibinfo {year} {1999})}\BibitemShut
  {NoStop}%
\bibitem [{\citenamefont {Carrillo-Tripp}\ \emph {et~al.}(2008)\citenamefont
  {Carrillo-Tripp}, \citenamefont {Shepherd}, \citenamefont {Borelli},
  \citenamefont {Venkataraman}, \citenamefont {Lander}, \citenamefont
  {Natarajan}, \citenamefont {Johnson}, \citenamefont {Brooks~III},\ and\
  \citenamefont {Reddy}}]{carrillo2008viperdb2}%
  \BibitemOpen
  \bibfield  {author} {\bibinfo {author} {\bibfnamefont {M.}~\bibnamefont
  {Carrillo-Tripp}}, \bibinfo {author} {\bibfnamefont {C.~M.}\ \bibnamefont
  {Shepherd}}, \bibinfo {author} {\bibfnamefont {I.~A.}\ \bibnamefont
  {Borelli}}, \bibinfo {author} {\bibfnamefont {S.}~\bibnamefont
  {Venkataraman}}, \bibinfo {author} {\bibfnamefont {G.}~\bibnamefont
  {Lander}}, \bibinfo {author} {\bibfnamefont {P.}~\bibnamefont {Natarajan}},
  \bibinfo {author} {\bibfnamefont {J.~E.}\ \bibnamefont {Johnson}}, \bibinfo
  {author} {\bibfnamefont {C.~L.}\ \bibnamefont {Brooks~III}}, \ and\ \bibinfo
  {author} {\bibfnamefont {V.~S.}\ \bibnamefont {Reddy}},\ }\href@noop {}
  {\bibfield  {journal} {\bibinfo  {journal} {Nucleic acids research}\ }\textbf
  {\bibinfo {volume} {37}},\ \bibinfo {pages} {D436} (\bibinfo {year}
  {2008})}\BibitemShut {NoStop}%
\bibitem [{\citenamefont {Hulo}\ \emph {et~al.}(2010)\citenamefont {Hulo},
  \citenamefont {De~Castro}, \citenamefont {Masson}, \citenamefont
  {Bougueleret}, \citenamefont {Bairoch}, \citenamefont {Xenarios},\ and\
  \citenamefont {Le~Mercier}}]{hulo2010viralzone}%
  \BibitemOpen
  \bibfield  {author} {\bibinfo {author} {\bibfnamefont {C.}~\bibnamefont
  {Hulo}}, \bibinfo {author} {\bibfnamefont {E.}~\bibnamefont {De~Castro}},
  \bibinfo {author} {\bibfnamefont {P.}~\bibnamefont {Masson}}, \bibinfo
  {author} {\bibfnamefont {L.}~\bibnamefont {Bougueleret}}, \bibinfo {author}
  {\bibfnamefont {A.}~\bibnamefont {Bairoch}}, \bibinfo {author} {\bibfnamefont
  {I.}~\bibnamefont {Xenarios}}, \ and\ \bibinfo {author} {\bibfnamefont
  {P.}~\bibnamefont {Le~Mercier}},\ }\href@noop {} {\bibfield  {journal}
  {\bibinfo  {journal} {Nucleic acids research}\ }\textbf {\bibinfo {volume}
  {39}},\ \bibinfo {pages} {D576} (\bibinfo {year} {2010})}\BibitemShut
  {NoStop}%
\bibitem [{\citenamefont {Nguyen}\ \emph {et~al.}(2006)\citenamefont {Nguyen},
  \citenamefont {Bruinsma},\ and\ \citenamefont
  {Gelbart}}]{nguyen2006continuum}%
  \BibitemOpen
  \bibfield  {author} {\bibinfo {author} {\bibfnamefont {T.}~\bibnamefont
  {Nguyen}}, \bibinfo {author} {\bibfnamefont {R.}~\bibnamefont {Bruinsma}}, \
  and\ \bibinfo {author} {\bibfnamefont {W.}~\bibnamefont {Gelbart}},\
  }\href@noop {} {\bibfield  {journal} {\bibinfo  {journal} {Physical review
  letters}\ }\textbf {\bibinfo {volume} {96}},\ \bibinfo {pages} {078102}
  (\bibinfo {year} {2006})}\BibitemShut {NoStop}%
\bibitem [{\citenamefont {Ning}\ \emph {et~al.}(2016)\citenamefont {Ning},
  \citenamefont {Erdemci-Tandogan}, \citenamefont {Yufenyuy}, \citenamefont
  {Wagner}, \citenamefont {Himes}, \citenamefont {Zhao}, \citenamefont {Aiken},
  \citenamefont {Zandi},\ and\ \citenamefont {Zhang}}]{nature2016}%
  \BibitemOpen
  \bibfield  {author} {\bibinfo {author} {\bibfnamefont {J.}~\bibnamefont
  {Ning}}, \bibinfo {author} {\bibfnamefont {G.}~\bibnamefont
  {Erdemci-Tandogan}}, \bibinfo {author} {\bibfnamefont {E.~L.}\ \bibnamefont
  {Yufenyuy}}, \bibinfo {author} {\bibfnamefont {J.}~\bibnamefont {Wagner}},
  \bibinfo {author} {\bibfnamefont {B.~A.}\ \bibnamefont {Himes}}, \bibinfo
  {author} {\bibfnamefont {G.}~\bibnamefont {Zhao}}, \bibinfo {author}
  {\bibfnamefont {C.}~\bibnamefont {Aiken}}, \bibinfo {author} {\bibfnamefont
  {R.}~\bibnamefont {Zandi}}, \ and\ \bibinfo {author} {\bibfnamefont
  {P.}~\bibnamefont {Zhang}},\ }\href {\doibase 10.1038/ncomms13689} {\bibfield
   {journal} {\bibinfo  {journal} {Nature Communications}\ }\textbf {\bibinfo
  {volume} {7}},\ \bibinfo {pages} {13689} (\bibinfo {year}
  {2016})}\BibitemShut {NoStop}%
\bibitem [{\citenamefont {Li}\ \emph {et~al.}(tted)\citenamefont {Li},
  \citenamefont {Roy}, \citenamefont {Travesset},\ and\ \citenamefont
  {Zandi}}]{siyu2018}%
  \BibitemOpen
  \bibfield  {author} {\bibinfo {author} {\bibfnamefont {S.}~\bibnamefont
  {Li}}, \bibinfo {author} {\bibfnamefont {P.}~\bibnamefont {Roy}}, \bibinfo
  {author} {\bibfnamefont {A.}~\bibnamefont {Travesset}}, \ and\ \bibinfo
  {author} {\bibfnamefont {R.}~\bibnamefont {Zandi}},\ }\href@noop {}
  {\bibfield  {journal} {\bibinfo  {journal} {Proceedings of the National
  Academy of Sciences}\ } (\bibinfo {year} {submitted})}\BibitemShut {NoStop}%
\bibitem [{\citenamefont {Lin}\ \emph {et~al.}(2012)\citenamefont {Lin},
  \citenamefont {van~der Schoot},\ and\ \citenamefont {Zandi}}]{Hsiang-Ku}%
  \BibitemOpen
  \bibfield  {author} {\bibinfo {author} {\bibfnamefont {H.-K.}\ \bibnamefont
  {Lin}}, \bibinfo {author} {\bibfnamefont {P.}~\bibnamefont {van~der Schoot}},
  \ and\ \bibinfo {author} {\bibfnamefont {R.}~\bibnamefont {Zandi}},\ }\href
  {\doibase 10.1088/1478-3975/9/6/066004} {\bibfield  {journal} {\bibinfo
  {journal} {Phys. Biol.}\ }\textbf {\bibinfo {volume} {9}},\ \bibinfo {pages}
  {066004} (\bibinfo {year} {2012})}\BibitemShut {NoStop}%
\bibitem [{\citenamefont {Kusters}\ \emph {et~al.}(2015)\citenamefont
  {Kusters}, \citenamefont {Lin}, \citenamefont {Zandi}, \citenamefont
  {Tsvetkova}, \citenamefont {Dragnea},\ and\ \citenamefont {van~der
  Schoot}}]{Kusters2015}%
  \BibitemOpen
  \bibfield  {author} {\bibinfo {author} {\bibfnamefont {R.}~\bibnamefont
  {Kusters}}, \bibinfo {author} {\bibfnamefont {H.-K.}\ \bibnamefont {Lin}},
  \bibinfo {author} {\bibfnamefont {R.}~\bibnamefont {Zandi}}, \bibinfo
  {author} {\bibfnamefont {I.}~\bibnamefont {Tsvetkova}}, \bibinfo {author}
  {\bibfnamefont {B.}~\bibnamefont {Dragnea}}, \ and\ \bibinfo {author}
  {\bibfnamefont {P.}~\bibnamefont {van~der Schoot}},\ }\href {\doibase
  10.1021/jp5108125} {\bibfield  {journal} {\bibinfo  {journal} {J. Phys. Chem.
  B}\ }\textbf {\bibinfo {volume} {119}},\ \bibinfo {pages} {1869} (\bibinfo
  {year} {2015})}\BibitemShut {NoStop}%
\bibitem [{\citenamefont {Moerman}\ \emph {et~al.}(2016)\citenamefont
  {Moerman}, \citenamefont {Van Der~Schoot},\ and\ \citenamefont
  {Kegel}}]{moerman2016kinetics}%
  \BibitemOpen
  \bibfield  {author} {\bibinfo {author} {\bibfnamefont {P.}~\bibnamefont
  {Moerman}}, \bibinfo {author} {\bibfnamefont {P.}~\bibnamefont {Van
  Der~Schoot}}, \ and\ \bibinfo {author} {\bibfnamefont {W.}~\bibnamefont
  {Kegel}},\ }\href@noop {} {\bibfield  {journal} {\bibinfo  {journal} {The
  Journal of Physical Chemistry B}\ }\textbf {\bibinfo {volume} {120}},\
  \bibinfo {pages} {6003} (\bibinfo {year} {2016})}\BibitemShut {NoStop}%
\bibitem [{\citenamefont {Wingfield}\ \emph {et~al.}(1995)\citenamefont
  {Wingfield}, \citenamefont {Stahl}, \citenamefont {Williams},\ and\
  \citenamefont {Steven}}]{wingfield1995hepatitis}%
  \BibitemOpen
  \bibfield  {author} {\bibinfo {author} {\bibfnamefont {P.~T.}\ \bibnamefont
  {Wingfield}}, \bibinfo {author} {\bibfnamefont {S.~J.}\ \bibnamefont
  {Stahl}}, \bibinfo {author} {\bibfnamefont {R.~W.}\ \bibnamefont {Williams}},
  \ and\ \bibinfo {author} {\bibfnamefont {A.~C.}\ \bibnamefont {Steven}},\
  }\href@noop {} {\bibfield  {journal} {\bibinfo  {journal} {Biochemistry}\
  }\textbf {\bibinfo {volume} {34}},\ \bibinfo {pages} {4919} (\bibinfo {year}
  {1995})}\BibitemShut {NoStop}%
\end{thebibliography}%



\end{document}